\RequirePackage{fix-cm}
\documentclass[smallextended]{svjour3}       
\smartqed  
\usepackage{graphicx}
\usepackage{cite}
\usepackage{amsmath,amssymb,amsfonts}
\usepackage{algorithmic}
\usepackage{hyperref}
\usepackage{graphicx}
\usepackage{textcomp}
\usepackage{tcolorbox}
\usepackage{multirow}
\usepackage{xcolor}
\usepackage{balance}
\usepackage[justification=centering]{caption}

\begin{document}

\title{Towards Speeding up Program Repair with Non-Autoregressive Model\\
}


\author{Zhenyu Yang  \and Yue Pan \and Zhen Yang \and    Zhongxing Yu
}


\institute{Zhenyu Yang \at
              72 Binhai Road, Jimo, Qingdao, P.R. China \\
              Shandong University \\
              \email{yangzycs@mail.sdu.edu.cn}           
           \and
            Yue Pan \at
              72 Binhai Road, Jimo, Qingdao, P.R. China \\
              Shandong University \\
              \email{pany@mail.sdu.edu.cn}           
           \and
           Zhen Yang \at
              72 Binhai Road, Jimo, Qingdao, P.R. China \\
             Shandong University \\
              \email{zhenyang@sdu.edu.cn}           
           \and
           Zhongxing Yu \at
              72 Binhai Road, Jimo, Qingdao, P.R. China \\
            Shandong University \\
              \email{zhongxing.yu@sdu.edu.cn}           
}

\date{Received: date / Accepted: date}

\maketitle

\begin{abstract}
Enlightened by the success of machine learning techniques in various application areas, recent years have witnessed a surge of research efforts on automatic program repair (APR) using machine learning techniques. Previous machine learning-based APR techniques essentially modified bugs in the autoregressive (AR) manner, which predicts future values based on past values. Due to the manner of token-by-token generation, the AR-based APR technique has a huge time delay. In particular, the delay of the APR model with a large number of parameters is more serious. The inability of fast repair negatively impacts the widespread adoption of machine learning-based APR techniques in real-life software development. To address the issue, we aim to apply the non-autoregressive (NAR) method to the APR task, which can output target code in a parallel manner to avoid huge repair delays. However, the naive use of the NAR manner for the APR task suffers from the issue of compromised patch quality. To effectively adapt the NAR manner for the APR task, we in this paper propose NARRepair, the first customized NAR code generation model for the APR task. The NARRepair model features three major novelties, including 1) the repair action predictor for alleviating the over-correction issue, 2) the inter-token dependency extractor for alleviating the issue of lacking inter-token dependency information, and 3) the two-stage decoder for alleviating the issue of lacking contextual information. We evaluated NARRepair on three widely used datasets in the APR community, and the results show that 1) compared to other APR techniques, the NARRepair model has the best performance within the limited repair time, and 2) compared to AR-based APR techniques, the repair speed of NARRepair has been increased by 1.4-6.4 times in the GPU environment. Overall, the results show that NARRepair has achieved state-of-the-art comprehensive performance in terms of repair speed and accuracy, highlighting the potential of the NAR model for speeding up program repair.

\end{abstract}

\section{Introduction}
\label{intro}
Program defects are inevitable during the software development process, and recent years have witnessed a surge of research efforts on automatic program repair (APR) to alleviate this issue \cite{le2011genprog, nguyen2013semfix, kim2013automatic, liu2018mining, le2016history,Prophet, rsrepair,angelix,10172854,yutse,yuemse}. Automatic program repair aims to automatically change the buggy code into correct code and promises to reduce software development costs and improve software reliability. To achieve this, different mechanisms have been explored in the ARP area, notably including search-based repair \cite{le2011genprog,yuan2018arja}, constraint-based repair \cite{nguyen2013semfix,mechtaev2015directfix}, and template-based repair \cite{kim2013automatic,liu2018mining}. Enlightened by the great success of machine learning in a wide variety of application areas, researchers have also investigated the use of machine learning for the APR task in the past few years \cite{yu2021deeprepair, chen2019sequencer, 9393494, zhu2021syntax, CODIT, lutellier2020coconut} and impressive results have been obtained. 
In particular, given that the autoregressive (AR) model has powerful inference capabilities, a majority of these {machine learning (ML)-based APR models} are basically built on top of the AR manner. For instance, SequenceR \cite{chen2019sequencer} is based on the AR manner and uses the copy mechanism to help the model repair buggy code. ThinkRepair \cite{yin2024thinkrepair} uses chain of thought to help large language models repair buggy code with the AR manner. 

The AR inference method predicts future values based on past values, \emph{i.e.}, it generates the target code sequence one by one. This use of AR manner leads to the inability of fast repair and huge time delays for repairing real-life complex bugs, which typically involves modifications to long code sequences. 
The longer it takes for the APR tools to fix bugs, the greater the potential losses caused by bugs. 
As reported by recent studies \cite{10.1145/3510003.3510040, 9609108, xiao2023expressapr,ahmed2020characterizing}, 
the response time of current APR tools greatly exceeds the patience of users and this time issue significantly limits the application scenario of APR approaches.
In the field of embedded systems (\emph{e.g.}, FPGA and Robots), the time issue is particularity of great relevance and several researchers have emphasized the importance of fast program repair \cite{gaimon1989real,steinbauer2005real, nazar2015improving}. 
Overall, the time issue overshadows the widespread adoption of ML-based APR techniques in real-life software development and calls for a general solution. 

\begin{figure}[]
\setlength{\belowcaptionskip}{-15pt}
\centering
\includegraphics[width=0.75\linewidth]{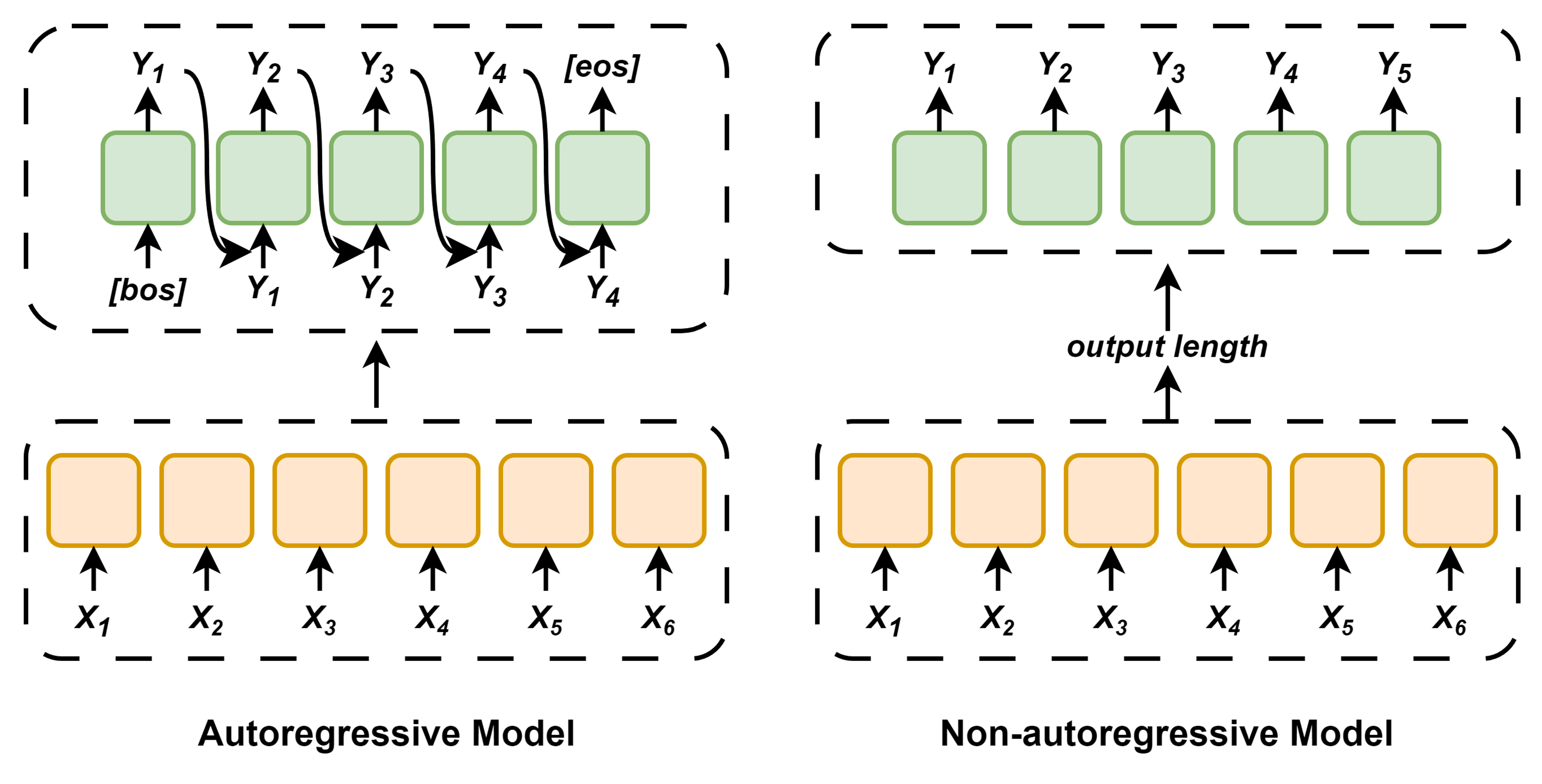}
\caption{\label{fig:frog1}The Inference Processes of the Autoregressive (AR) Model and the Non-autoregressive (NAR) Model.}
\end{figure}

To address this issue, inspired by the work on non-autoregressive translation (NAT)\cite{gu2018non, li2019hint,10.1145/3649594}, we propose to design techniques on top of the non-autoregressive (NAR) model. Unlike the AR model, the NAR model \cite{gu2018non} generates the target code in parallel and thus can greatly improve the speed of model inference. The inference processes of the AR model and the NAR model are illustrated in Figure \ref{fig:frog1}. 

While the NAR models have been widely used in the field of machine translation (MT), naive use of the NAR models in the MT field for the APR task results in poor performance according to the results of our pre-experiments. In particular, there are three major limitations. First, \textit{over-correction issue can arise}. The process of MT involves converting text in one language into another language, so the NAR models of MT will modify each token in the source text into a token in other languages. However, unlike MT, the program tokens that need modification account for only a small portion of the buggy code in the APR task. It is difficult for the NAR model to identify all the correct program tokens in the buggy code and keep them without making any modifications. Thus, the NAR model will suffer from the over-correction issue, \emph{i.e.}, many correct program tokens can instead be modified into wrong ones. Second, 
\textit{the issue of lacking inter-token dependency information can arise.} Unlike the AR model which generates tokens sequentially, the NAR model generates tokens in parallel. In other words, the NAR model does not know what tokens are elsewhere when it generates a token. Thus, the NAR model necessarily loses the inter-token dependency information that is vital for sentence grammar correctness. 
In particular, given that programs feature formal grammar and rich structural dependency, losing inter-token dependency information will significantly deteriorate the performance of the APR task. 
Finally, \textit{the issue of lacking contextual
information can arise}. The effectiveness of the {ML-based APR models} heavily relies on using the contextual information to modify wrong program tokens. However, since the NAR model generates tokens in parallel and processes context and wrong tokens alike, it cannot effectively obtain the contextual information. Hence, the repair performance of the model will be compromised.

To overcome the above limitations, we in this paper propose a novel NAR model, named NARRepair, for generating fixed code quickly while ensuring accuracy. For the architecture, similar to most NAR models, NARRepair uses the encoder-decoder structure on top of the transformer model. 

NARRepair features three major novelties for addressing the three limitations in order. First, \textit{NARRepair uses the repair action predictor for alleviating the over-correction issue}. 
When generating a target token, the NAR model generates the probabilities of all tokens in the dictionary and selects the one with the highest probability. 
However, a dictionary that contains tens of thousands of tokens will inevitably have noise.
To address this issue, we propose an action-guided repair method that divides the repair processes into four actions: ``keep'', ``replace'', ``insert'', and ``delete''. Instead of directly predicting the correct program token, NARRepair predicts the repair action for each token in the buggy code. 
This method noticeably reduces the complexity of the probability distribution that the model needs to learn.
Second, \textit{NARRepair uses the inter-token dependency extractor for alleviating the issue of lacking inter-token dependency information}.
It is reasonable to assume that a node in the abstract syntax tree (AST) has an association with its parent node(s) and has the strongest association with the nearest parent node. Furthermore, the nearest common parent node of a token pair has a strong association with both tokens and can represent the dependency information between the two tokens. Thus, to enable the NAR model to obtain dependency information between tokens (especially dependency information at the semantic level), 
we propose an inter-token dependency extractor that learns the dependency information between token pairs through the nearest common parent nodes in the AST. 
Finally, \textit{NARRepair uses the two-stage decoder for alleviating the issue of lacking contextual information}. To recover contextual information, we propose a decoding method that divides the decoding process into two stages. The first stage generates a preliminary result of the fixed program tokens. We keep the tokens with high confidence, including (1) that the repair action of the generated token is consistent with the previously predicted repair action, and (2) that whose prediction probabilities are greater than the threshold. These kept high-confidence tokens are used as context for the remaining low-confidence tokens that are masked, and the purpose of the second
stage is to regenerate the masked tokens based on the contextual information.






To evaluate the performance of NARRrepair, we assess its {repair} speed and accuracy on three widely used datasets for the APR task: Defect4J v1.2 \cite{just2014defects4j} that contains 395 bugs, Defect4J v2.0 that contains 420 bugs, and QuixBugs \cite{lin2017quixbugs} that contains 40 bugs. In terms of accuracy, NARRepair fixes more bugs than all APR baseline models within the time limits. For the defects4j v1.2 dataset, NARRepair fixes 9, 15, and 10 more bugs than the best baseline under the 3-minutes, 5-minutes, and 10-minutes time limits, respectively. For the defects4j v2.0 dataset, NARRepair fixes 7, 11, and 5 more bugs than the best baseline under the 3-minutes, 5-minutes, and 10-minutes time limits, respectively. For the QuixBugs dataset, NARRepair fixes 4, 5, and 3 more bugs than the best baseline under the 3-minutes, 5-minutes, and 10-minutes time limits, respectively. In terms of {repair} speed, compared with other autoregressive APR models, the {repair} speed of NARRepair is 1.4-6.4 times faster in the GPU environment. These results show that the proposed NARRepair model has the best comprehensive performance in terms of {repair} speed and accuracy. 

In summary, our contributions in this work are as follows:
\begin{itemize}
    \item We propose the NARRepair model for the APR task. To the best of our knowledge, NARRepair is the first NAR model designed for the APR task. 
    \item We propose three techniques to overcome the limitations of the naive use of NAR model for the APR task, including 1) the repair action predictor for alleviating the over-correction issue, 2) the inter-token dependency extractor for alleviating the issue of lacking inter-token dependency information, and 3) the two-stage decoder for alleviating the issue of lacking contextual information.
    \item We evaluated the NARRepair model on three widely used datasets for the APR task, and the results show that the NARrepair model has the best comprehensive performance in terms of {repair} speed and accuracy.
\end{itemize}

The remainder of this paper is structured as follows. We
first give closely related work in Section~\ref{relatedwork}, followed by Section~\ref{method} which describes the NARRrepair approach in detail. Section~\ref{evaluationsetup} and 
Section~\ref{evaluationresult} present the evaluation setup and evaluation results, respectively. Finally, Section~\ref{conclusion} concludes the paper and gives future perspectives. This paper is a major revision of an Arxiv preprint \cite{yangoldNARRepair}. 

 Our replication package is publicly available at \url{https://github.com/mlyzy/Speed_Repair}.

\section{Related Work}\label{relatedwork}
This section reviews some work closely related with this article, including automatic program repair, machine learning-based repair, and non-autoregressive models.

\subsection{Automatic Program Repair} Given the time-consuming and error-prone nature of program debugging \cite{debugging,multiple-fault,10.1145/3611643.3616338,urli2018design,yuguifl,10.1016/j.infsof.2013.07.004}, automatic program repair (APR) techniques have been proposed to reduce software development costs and improve software reliability. Recent years have witnessed a surge of APR techniques rooted in different disciplines, notably including search-based repair, constraint-based repair, and template-based repair. Search-based APR techniques \cite{le2011genprog,yuan2018arja, jobstmann2005program}
treat generating patches as finding feasible solutions in a predefined search space. Constraint-based APR techniques \cite{xuan2016nopol,nguyen2013semfix,wei2010automated} guide the repair process by first developing a set of constraints and then solving these constraints to derive the patches. Template-based APR techniques \cite{kim2013automatic,le2016history,liu2019tbar,long2015staged} rely on various targeted repair templates to generate patches, which have good repair effects for specific bug types.

\subsection{Machine Learning-based Repair} 
Enlightened by the huge success of machine learning in various application areas, researchers have also investigated the use of machine learning for the APR task in recent years. As a result, there are an abundance of {ML-based APR models} in the literature. 
Gupta et al. \cite{gupta2017deepfix} propose DeepFix, an APR model to repair C compilation defects. DeepFix is a multi-layered sequence-to-sequence neural network that directly locates and fixes defects. 
White et al. \cite{yu2021deeprepair} propose an APR model named DeepRepair, which infers code similarity through machine learning and can sort code fragments based on their similarities with suspicious elements.  
Chen et al. \cite{chen2019sequencer} propose a technique named SequenceR for end-to-end APR on top of the sequence-to-sequence model, which uses abstract context to simulate the process of analyzing and fixing bugs conducted by developers. Lutellier et al. \cite{lutellier2020coconut} propose CoCoNuT, a technique for APR using a neural machine translation model based on the CNNs. Zhu et al. \cite{zhu2021syntax} propose Recoder, which constrains the output of the APR model via syntax rules to repair fine-grained defects. 
Ye et al. \cite{ye2022neural} propose RewardRepair, which adds test information to the model to ensure that candidate patches are compilable. 
Xia et al.\cite{xia2022less} treat program repair tasks as text fill-in-the-blanks and generate patches based on the contextual information. Meng et al. \cite{meng2023template} convert the repair task into a cloze task through a template and use a pre-trained model to generate patches. 
Xia et al. \cite{xia2024automated} proposed a dialogue-driven APR method by combining patch generation with feedback information. 
Yin et al. \cite{yin2024thinkrepair} use the Chain-of-Thought to guide large language models to fix bugs. 

At present, a majority of these {ML-based APR models} basically generate correct code in the AR manner. The AR model requires that the output of each step waits for the output of the previous position in order, resulting in slow reasoning. Consequently, this use of AR manner leads to the inability of fast repair and huge time delays for repairing real-life complex bugs, which typically involves modifications to long code snippets. These negative consequences create obstacles to the adoption of {ML-based APR models} in real-life software development and maintenance \cite{10.1145/3510003.3510040, 9609108, xiao2023expressapr,ahmed2020characterizing}.

\subsection{Non-autoregressive Models} The purpose of the non-autoregressive (NAR) models is to reduce inference time by generating target sentences in parallel. Gu et al. \cite{gu2018non} propose the first NAR model, which assumes that all tokens in the target sentence are independent of each other and can output all target tokens in parallel in one step. Shu et al. \cite{shu2020latent} use a spherical Gaussian to generate latent variables for each input token to increase the dependence between tokens in the target sentence. Ran et al. \cite{ran2021guiding} use latent variables to establish the position information of the target token. 
Ma et al. \cite{ma2019flowseq} use generative flow to model latent variables containing rich information.
Stern et al. \cite{stern2019insertion} propose a NAR model based on insertion operations, which generates a subsequence of the final result sequence through iteration at each step until all insertion operations are empty.
Gui et al. \cite{gui2023non} use probabilistic context-free grammar to enhance the ability of NAT model to capture complex dependencies between output tokens.
 {Bao et al. \cite{bao2023non} apply the NAR inference method to the document translation task and achieve significant acceleration. Liu et al.\cite{liu2023selective} use selective knowledge distillation to improve the NAR model training effect. Tan et al. \cite{tandiffnorm} simplify the data distribution through a diffusion-based normalization strategy to reduce the impact of noise on the NAR model.}
However, given the three major limitations outlined in Section~\ref{intro}, naively using existing NAR models for the APR task cannot obtain satisfactory results. 
Therefore, we propose the NARRepair model in this paper to meet the unique needs of the APR task.
\begin{figure*}[]
\setlength{\belowcaptionskip}{-15pt}
\centering
\includegraphics[width=0.95\textwidth]{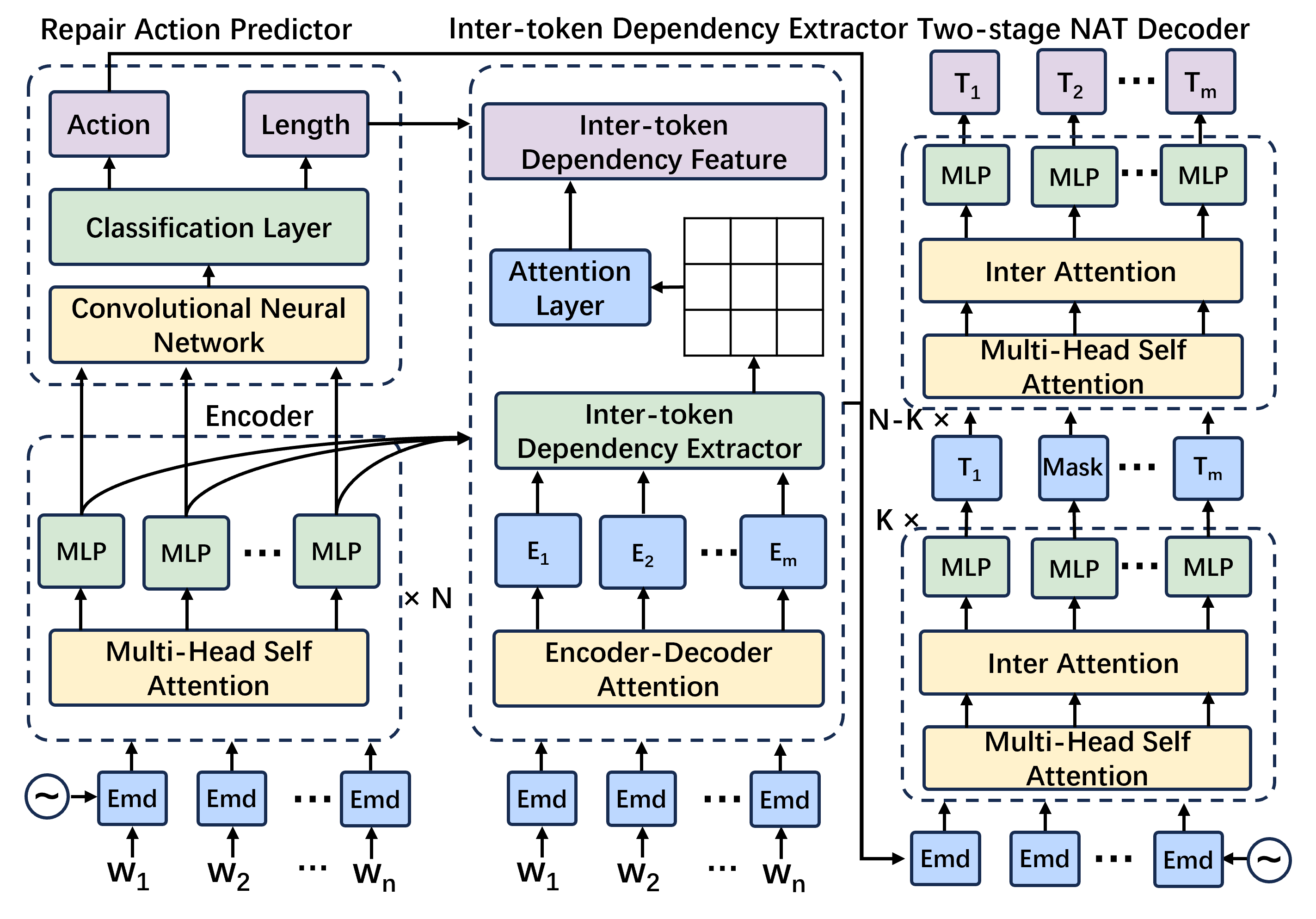}
\caption{\label{fig:frog2}An Overview of the NARRepair Architecture. }
\end{figure*}

\section{NARRepair}\label{method}
This section introduces the NARRepair model, which uses the NAR manner to generate program tokens in parallel to improve the {repair} speed. The model consists of four parts: code encoder, repair action predictor, inter-token dependency extractor, and two-stage NAR decoder. Figure \ref{fig:frog2} shows the structure of the NARRepair model. NARRepair proceeds as follows: 
\begin{itemize}
    \item Code encoder embeds the buggy code into feature vectors (§3.1).
    \item Given the buggy code feature, the repair action predictor predicts repair action and output length for each word (§3.2).
    \item According to the output length, the inter-token dependency extractor first generates the feature vector of the repaired code text. Then, the extractor obtains the inter-token dependency information and fuses it with the word feature vector to obtain the word feature vector with dependency information (§3.3).
    \item Given the word feature vector from the previous step, the two-stage decoder generates all repaired words (§3.4).
\end{itemize}
The following sections elaborate on the process.

\subsection{Code Encoder}
The code encoder can extract features from the buggy code text ${W}_{i:n}$ and convert ${W}_{i:n}$ into a token embedding ${E}_{i:n}$. The output token embedding ${E}_{i:n}$ can be used
for the prediction and tagging of subsequent modules. We use the encoder part of the transformer model \cite{vaswani2017attention} as the code encoder of the model.

Here, we briefly give an overview of the encoder of the transformer model. The encoder of the transformer is composed of multiple identical layers, and each layer has two sub-layers. The first is a multi-head self-attention layer that fuses token features by calculating the attention weight between token feature vectors. The second sub-layer is a feedforward neural network used to normalize the output of the model. Residual links are used between sub-layers. The operation process of the code encoder can be defined as
\begin{equation}\label{eq1}
{E}_{i:n}=Encoder({W}_{i:n}+{W}_{pos})
\end{equation}
and the operation of each encoder layer can be expressed as
\begin{align}\label{eq2}
{X}^{l}_{attention}&={X}^{l-1}_{hidden}+Attention({X}^{l-1}_{hidden})\\
{X}^{l}_{hidden}&=Feedforward({X}^{l}_{attention})\\
{X}^{0}_{hidden}&={W}_{i:n}+{W}_{pos}
\end{align}
where ${X}_{pos}$ is position embedding, $Attention$ is self-attention layer, and $Feedforward$ is feedforward neural network layer.

\subsection{Repair Action Predictor}\label{action}
\begin{figure}
\setlength{\belowcaptionskip}{-15pt}
\centering
\includegraphics[width=0.6\linewidth]{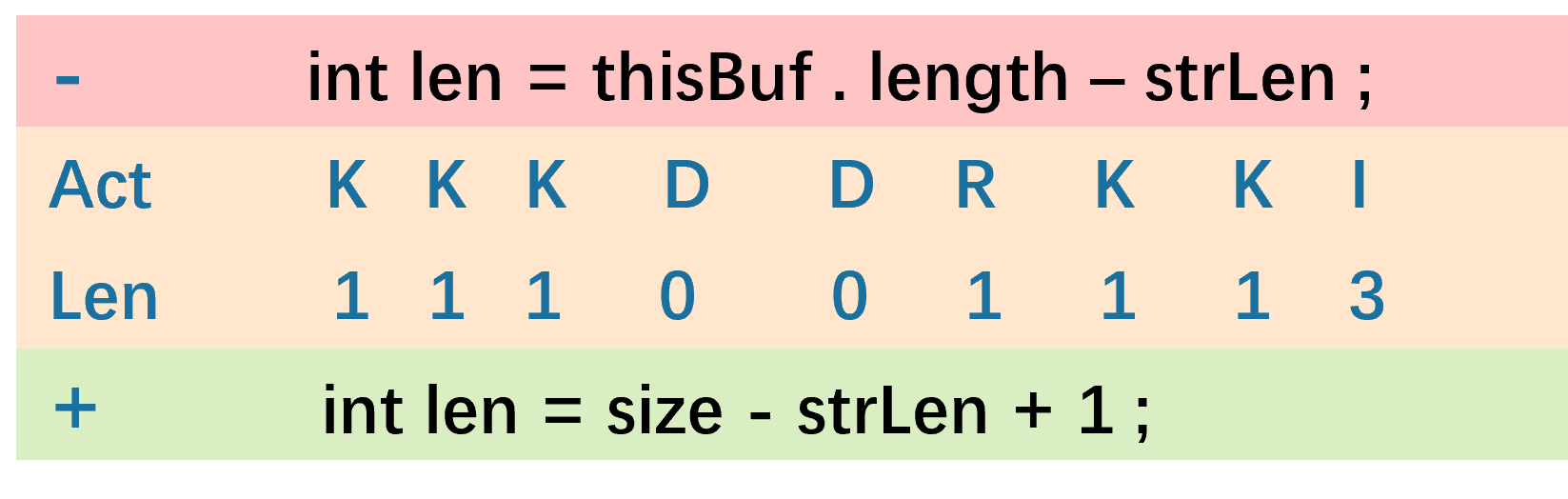}
\caption{\label{fig:frog3} An Example of Repair Action Prediction for a Real Bug. }
\end{figure}

The repair action predictor predicts the repair action for each token in the buggy code. The content of the repair action is divided into two parts: the type of the repair action and the repair length. We classify all repair actions into 4 categories: ``keep'', ``insert'', ``delete'', and ``replace''. The repair length represents the number of generated repair tokens for each fixed token. Typically, the repair length for actions ``replace'' and ``keep'' is 1; the repair length for action ``delete'' is 0; the repair length for action ``insert'' is the number of tokens inserted.
Figure \ref{fig:frog3} gives an example of repair action prediction for the ``Lang-61'' bug in the Defect4j dataset. Compared to NAR models in machine translation that need to predict the probabilities of all dictionary tokens, the repair action predictor only needs to predict the probabilities of four repair actions. When the predicted action is ``keep'', the model does not need to change the tokens. This method effectively alleviates the issue of modifying the correct tokens into the wrong ones. 

Regarding the model structure, since convolutional neural networks have the advantage of effectively acquiring local features, the repair action predictor uses a convolutional neural network \cite{kalchbrenner2014convolutional} to extract the token features after receiving the output of the encoder. Then, the classification layer predicts the repair action and the repair length for each token in the buggy code text separately. The detailed operations are as follows:
\begin{align}\label{eq2}
{X}_{feature}&=ConV({E}_{1:n})\\
{Act}_{1:n}={Linear}&_{1}(Dropout(Relu({X}_{feature})))\\
{Len}_{1:n}={Linear}&_{2}(Dropout(Relu({X}_{feature})))
\end{align}
where $ConV$ is the convolutional neural network layer, ${Linear}_{1}$ is a fully connected layer whose output dimension is the number of repair actions, and ${Linear}_{2}$ is a fully connected layer whose output dimension is the maximum length. We use the cross-entropy method to compute the loss between the predictor output and the ground truth as:\\
\begin{align}\label{eq2}
{L}_{lenth}&=-\sum ^{n}_{i} {log{p}_{lenth}({l}_{i}|X,\, M)}\\
{L}_{act}&=-\sum ^{n}_{i} {log{p}_{act}({a}_{i}|X,\, N)}
\end{align}
where $M$ and $N$ are model parameters, $X$ is the input token feature vector, ${l}_{i}$ is the repair length of the i-th token, and ${a}_{i}$ is the repair action of the i-th token.
\begin{figure*}
\setlength{\belowcaptionskip}{-15pt}
\centering
\includegraphics[width=1\textwidth]{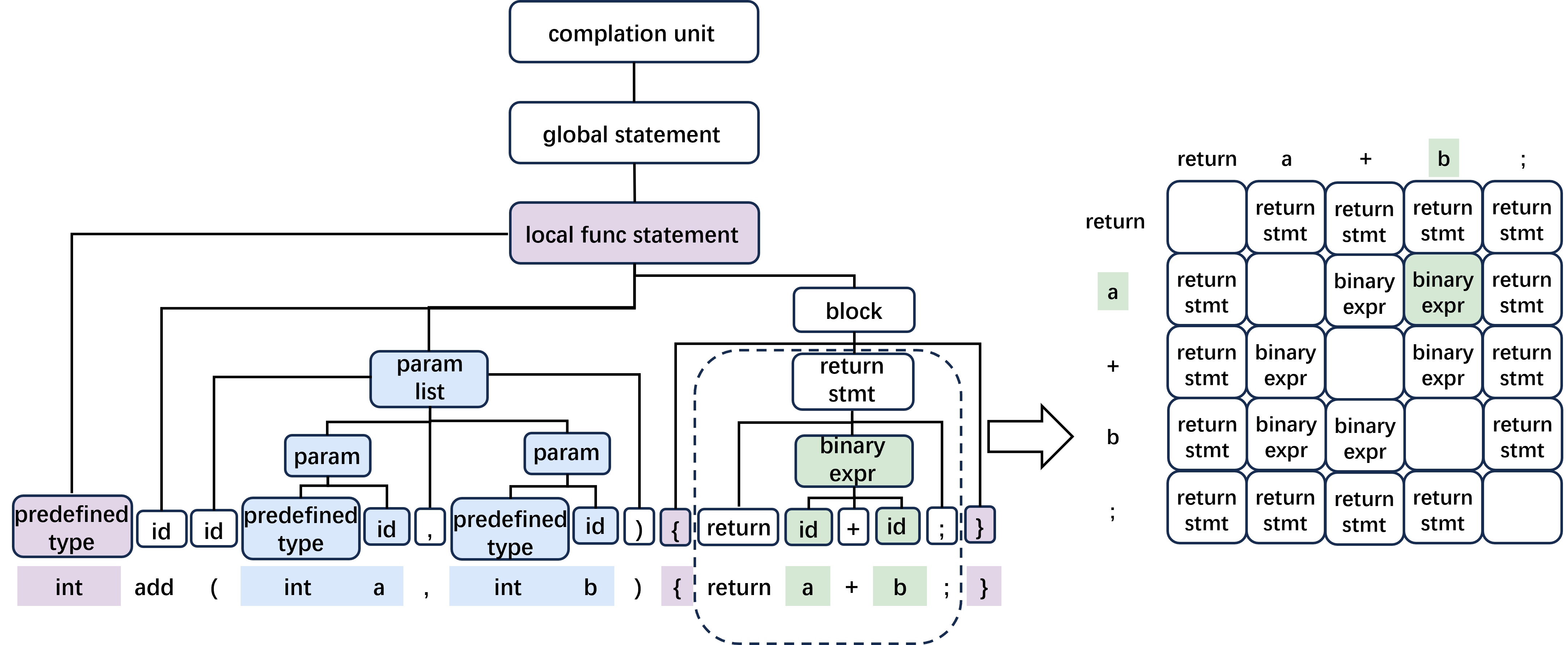}
\caption{\label{fig:frog4}The Example of Generating AST and Inter-token Dependency Matrix for the Code \texttt{return a+b;}.}
\end{figure*}

\subsection{Inter-token Dependency Extractor}    
The inter-token dependency extractor can learn the dependency information between token pairs through the nearest common parent nodes in the AST. To get the nearest parent node of token pairs, we need to generate the AST for the code text. Given the code text, we use the program analysis tool Tree-sitter\cite{tree-sitter} to extract its AST. Figure \ref{fig:frog4} shows the code text \texttt{return a+b;} and its corresponding AST. In this example, the nearest common parent of \texttt{a} and \texttt{b} on the right is \texttt{binary expr}. We list the nearest common parent nodes of all token pairs and generate a token dependency matrix. We show the inter-token dependency matrix of \texttt{return a+b;} in the right part of Figure \ref{fig:frog4}. The relationship between token \texttt{{a}} and token \texttt{{b}} is equivalent to that between \texttt{{a}} and \texttt{{b}}, so the dependency matrix is symmetric. We use the obtained dependency matrix as the ground truth to train the inter-token dependency extractor. 

\begin{figure}
\centering
\includegraphics[width=0.6\linewidth]{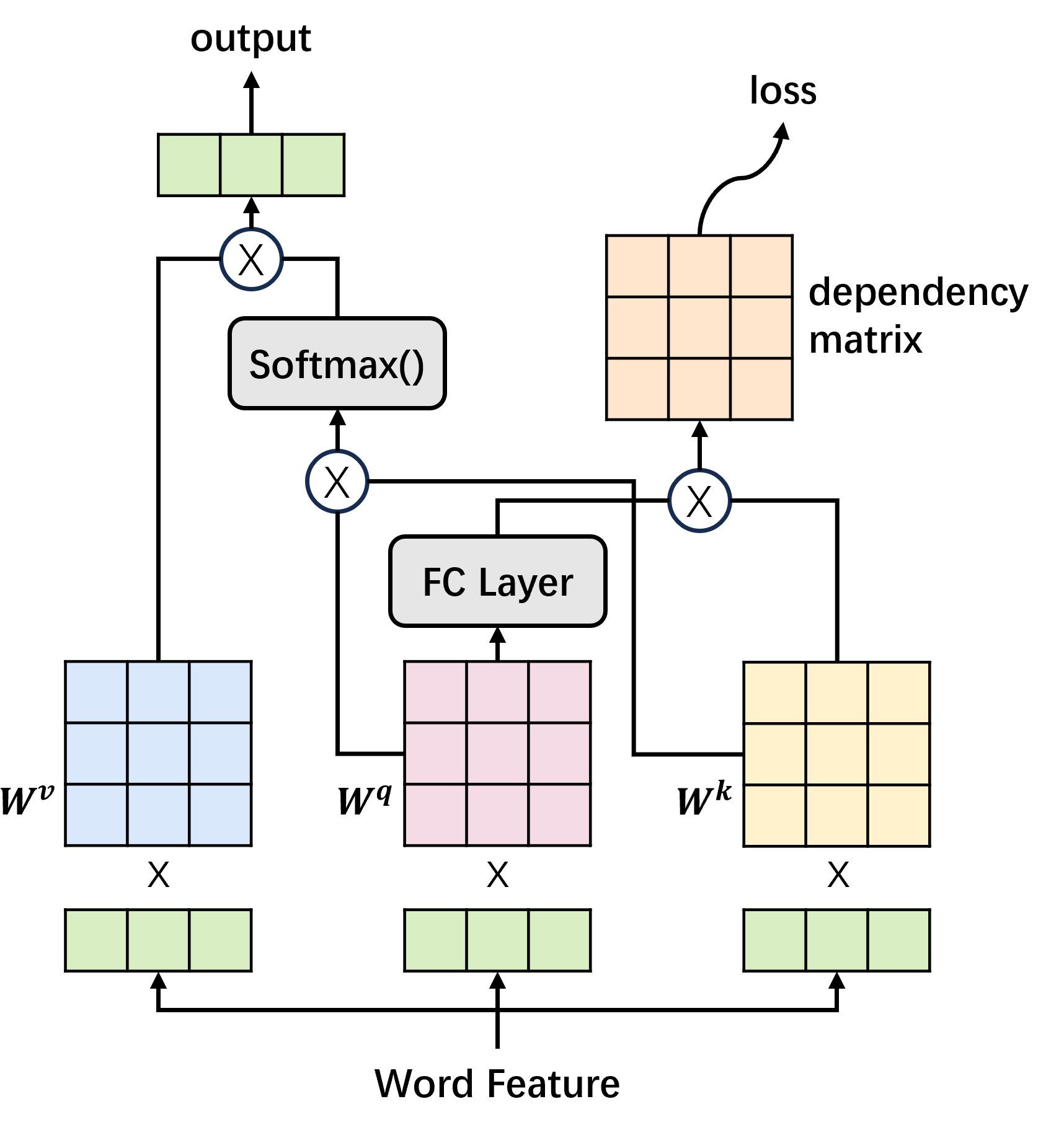}
\caption{\label{fig:frog5}The Structure of the Inter-token Dependency Extractor.}
\end{figure}

Given the feature vector {${E}_{1:n}$} of the faulty text, we first use the encoder-decoder attention module to obtain the feature vector ${D}_{1:m}$ of the target text. Then, we use the inter-token dependency extractor to predict the nearest common parent nodes of token pairs as dependency information. Inspired by the work in \cite{dozat2016deep}, the inter-token dependency extractor uses an attention mechanism to model dependency information. For each target text feature, the extractor maps the feature using the query ${W}^{q}$ and the key ${W}^{k}$. The dot product of maps Q and K predicts the token dependency matrix. To facilitate training, we construct an index table of parent nodes in the AST and replace the parent nodes in the matrix with index values as ground truth. Figure \ref{fig:frog5} shows the structure of the inter-token dependency extractor. The detailed operations are as follows:
\begin{align}\label{eq2}
Q&={W}^{q}*{D}_{1:m}\\
K&={W}^{k}*{D}_{1:m}\\
M_{dependency}&=Linear3(Q)*K^{T}
\end{align}
where ${W}_{q}$ and ${W}_{k}$ are the weight matrix, and $Linear3$ is a fully connected layer used to convert the feature dimension into the dimension of the number of parent nodes in the AST. After obtaining the dependency information, we use an attention machine to fuse it into token features in order for the decoder to obtain the dependency information about other tokens when generating tokens. The operations are as follows:
\begin{align}\label{eq2}
score &= softmax(\frac {Q{K}^{T}} {\sqrt {d}})\\
V&={W}^{v}*{D}_{1:m}\\
{H}_{1:m} &= \sum ^{n}_{i=1} {score*V}
\end{align}
where ${W}^{v}$ is the weight matrix, $softmax$ is the normalization function for calculating fractions. We compute the loss of the inter-token dependency extractor as:\\
\begin{align}\label{eq2}
{L}_{depend}=-log(M|D,E)
\end{align}
where D is the input feature vector, E is the parameter of the extractor, and M is the output token dependency matrix.

\subsection{Two-stage NAR Decoder}\label{decoder}
The two-stage NAR decoder decomposes the normal NAR decoder into two parts for step-by-step decoding. 
The purpose of the first stage of decoding is to generate a preliminary result and retain tokens with high confidence. For the text generated by the first stage of decoding, we retain two parts of the tokens whose confidence levels are high: (1) the repair action of the generated token is consistent with the previously predicted repair action and (2) tokens whose prediction probabilities are greater than a threshold.

We set the threshold to be 0.7 in our experimental evaluation. We retain these high-confidence tokens and use them as context for the remaining low-confidence tokens. Since the remaining tokens may be wrong, the purpose of the second decoding is to regenerate the tokens with low confidence based on the contextual information. We mask the remaining tokens with the \texttt{[Mask]} tag. Literature about mask language models \cite{devlin2018bert,feng2020codebert,liu2019roberta,guo2020graphcodebert} has proven that the attention mechanism can obtain contextual information for tokens with the \texttt{[Mask]} tag. Hence, the result of the second stage of decoding based on contextual information will be more accurate. 

\begin{figure}
\centering
\includegraphics[width=0.50\linewidth]{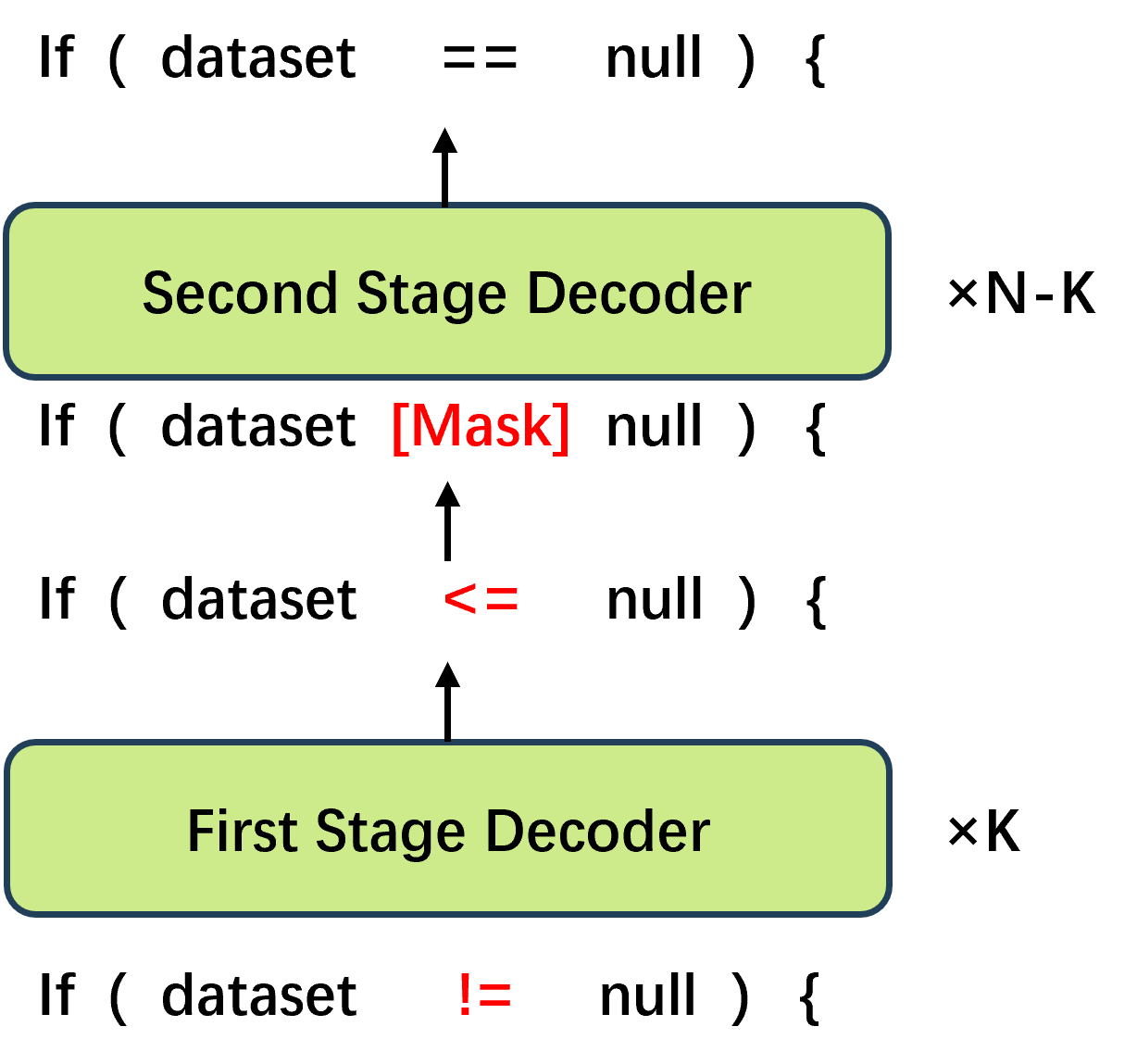}
\captionsetup{position=top}
\caption{\label{fig:frog6}The Decoding Process of the Two-stage NAR Decoder.}
\end{figure}

Take the code text \texttt{if ( dataset != null ) \{} as an example, Figure \ref{fig:frog6} shows the decoding process of the two-stage NAR decoder. First, the first stage decoder generates the preliminary result \texttt{if ( dataset \textless= null ) \{} and considers \texttt{\textless=} as having low confidence. Then, the second stage decoder masks the \texttt{\textless=} with the \texttt{[Mask]} tag to obtain the context information and generate the correct result.
Assuming that the given input feature is ${H}_{1:m}$, the specific operation of the first stage decoder is as follows:
\begin{align}\label{eq2}
{E}_{first}={Decoder}_{first}&({H}_{1:m})={Layer}^{1:n-k}_{decoder}({H}_{1:m})
\end{align}
where ${Layer}^{1: n-k}_{decoder}$ is the $1$st to $n-k$th layer of the decoder and ${E}_{first}$ is the feature vector outputted by the first stage of decoding. 
We further mask the result of the first stage of decoding and input it into the second stage decoder as follows:
\begin{align}\label{eq2}
{E}_{first}={Decoder}_{first}&({H}_{1:m})={Layer}^{1:n-k}_{decoder}({H}_{1:m})\\
{P}_{first}&=Softmax({E}_{first})
\end{align}
where ${Layer}^{1: n-k}_{decoder}$ is the 1st to $n-k$th layer of the decoder, ${E}_{first}$ is the feature vector outputted by the first stage of decoding, and ${P}_{first}$ is the probability distribution of the first stage of decoding. 
We further mask the result of the first stage of decoding and input it into the second stage decoder as follows:
\begin{align}\label{eq2}
{E}_{mask}&= MaskFunc({E}_{first})\\
{E}_{second}={Decoder}_{second}&({E}_{mask})={Layer}^{n-k:n}_{decoder}({E}_{mask})\\
{P}_{second}&=Softmax({E}_{second})
\end{align}
where $MaskFunc$ is the masking function used to replace tokens with low confidence with the \texttt{[MASK]} tag, ${Layer}^{n-k : n}_{decoder}$ is the $n-k$th to $n$th layer of the decoder, ${E}_{second}$ is the feature vector outputted by the second stage of decoding, and ${P}_{second}$ is the token probability distribution generated by the second stage of decoding. 
We compute the loss based on the output of the decoder as:
\begin{align}\label{eq2}
{{L}_{dec}}_{}&=-\sum ^{m}_{i} {log({P}_{dec}(R|H,\partial ))}
\end{align}
where $R$ is the output result of the decoder, $H$ is the hidden feature, and $\partial$ is the decoder parameters.

\subsection{Training and Inference}
During the training process, the three sub-modules of the NARRepair model are jointly learned. The final loss function of NARRepair is as follows:
\begin{align}\label{eq2}
L={L}_{dec}+\alpha({L}_{act}+{L}_{length})+\lambda{L}_{depend}
\end{align}
where $\alpha$ and $\lambda$ are the hyperparameters used to adjust the importance of each training loss. We set both $\alpha$ and $\lambda$ to be 0.1 in our study.

During inference, the repair action predictor firstly predicts the repair action and output length for each token in the buggy code. Then, based on the output length, the encoder-decoder attention layer transforms the buggy code's feature vector into the fixed code's feature vector. Next, the inter-token dependency extractor generates fixed code feature vectors with inter-token dependency information through AST. Finally, the two-stage decoder outputs the fixed code based on the repair action and the fixed code feature vector with inter-token dependency information. For each buggy code, we generate 200 patches and test whether they contain the correct patches.

After obtaining the inference results, we use the ExpressAPR tool proposed by Xiao et al. \cite{xiao2023expressapr} to verify whether the patch is correct. ExpressAPR includes five existing acceleration techniques for mutation testing \cite{mutationtesting,article,compileroptimization}, including mutant schemata, mutant deduplication, test virtualization, test prioritization, and parallelization. ExpressAPR can improve the speed of patch verification while ensuring the accuracy of verification results.

\section{Evaluation Setup} \label{evaluationsetup}
To prove the validity of our idea, we train the NARRepair model and evaluate its performance. In this section, we introduce the evaluation setup.

\subsection{Research Questions}
This paper explores the following research questions:  

\noindent
\textbf{RQ1: Compared to other APR models within time limits, what is the performance of NARRepair?} This is an overall question about the performance of the NARRepair model. For this question, to simulate the urgency of fixing bugs, we set time limits for fixing bugs. We evaluate the NARRepair model and compare its performance with that of other APR models within time limits.

\noindent
\textbf{RQ2: How does each module contribute to the final result of NARRepair?} For this question, we gradually remove submodules from the complete NARRepair model and investigate the contribution of each submodule.

\noindent
\textbf{RQ3: Can NARRepair predict repair action and repair length?} The repair action and repair length directly affect the final output of NARRepair. This problem aims to analyze whether NARRepair can accurately predict the repair action and repair length of the bugs.

\noindent
\textbf{RQ4: Can the NARRepair model effectively alleviate the over-correction problem?} One of the important limitations we aim at is the over-correction issue. To study the effectiveness of our idea for alleviating this issue in detail, we compare the number of correct tokens that NARRepair changed into incorrect tokens before and after removing the repair action predictor module and the two-stage decoder module. 

\noindent
\textbf{RQ5: Whether the nearest parent node is closely related to its child node in the AST?} We assume that the nearest parent node has a close relationship with its child nodes. This question aims to investigate the validity of this assumption. 

\subsection{Data Collection}
We use the dataset published by selfAPR\cite{ye2022selfapr} on GitHub as our model dataset, which contains code text pairs (buggy and correct) generated for common types of program errors by self-supervision. We discard some data for which we could not generate the AST and retain 837, 059 pieces of data. To prevent data leakage, we remove projects related to Defect4J\cite{just2014defects4j} and QuixBugs\cite{lin2017quixbugs}. Finally, our dataset contains 1.52 million instances, and we use 90\% of the dataset as training data and 10\% as validation data.

We use three widely used datasets for the APR task to evaluate the model performance. The first one is Defect4J v1.2 \cite{just2014defects4j}, which contains 395 bugs in real Java projects. The second one is Defect4J v2.0 \cite{just2014defects4j}, which contains 438 additional bugs compared to Defect4J v1.2. The third one is QuixBugs \cite{lin2017quixbugs}, which contains 40 Java bugs and is introduced to test the model robustness on data distributions besides the Defect4J bugs.

\subsection{Data Preprocessing}
For training data, we need to generate repair action(s) and inter-token dependency matrix for each piece of data to facilitate model learning. For generating repair actions, we use dynamic programming to calculate the edit distance between the buggy code text and the fixed code text. After calculating the edit distance, we use backtracking to output the specific repair action for each step. For example, for a code pair containing the buggy code text \texttt{if ( result != null )} and the fixed code text \texttt{if ( ! result . isNotype ( ) )}, we can accordingly get the required repair actions on top of this procedure: Keep, Keep, Insert, Replace, Replace, Insert. For the inter-token dependency matrix, we first use the Tree-sitter tool \cite{tree-sitter} to generate an AST for each correct code piece and obtain the path from each token to the root node. Then, we compare the paths of different tokens and add the nearest common parent of the token pair to the dependency matrix. We establish the repair action(s) and inter-token dependency matrix for each piece of training data and feed the established data to the model. 

\subsection{Knowledge Distillation}
Due to the conditional independence assumption of the NAR model, it is difficult to capture the multimodal distribution of target text, which is called the ``multi-modality problem'' in the literature\cite{gu2018non}. For example, the buggy code \texttt{Node block = NodeUtil.getFunctionBody ( fnNode );} corresponds to two fixed code pieces \texttt{Node argsNode = NodeUtil.getFnParameters ( fnNode );} and \texttt{Node block = fnNode.getLastChild ( );}. Since the NAR model outputs results in a parallel manner and lacks information about other locations, for the above example, the NAR model may output \texttt{argsNode} in the second location and \texttt{fnNode.getLastChild} in the fourth location. While these outputs all correspond to the correct code, they do not combine correctly. This situation is very common in the dataset and seriously affects the performance of the NAR model.

To alleviate this problem, we refer to previous NAR works \cite{gu2018non,wang2019non,qian2020glancing} and use the knowledge distillation method to process the training dataset. First, we train the CodeT5-large pre-trained model with the training dataset. Then, we use the text generated by the trained CodeT5-large on the original training dataset as the distilled training dataset. This kind of processing can remove the noise in the dataset as much as possible, and only retain the most correctly fixed code. The model trained on the distilled training dataset can learn the knowledge of program repair more accurately. To account for the model's robustness, we train the NARRepair model on the distilled and original training dataset simultaneously and learn primarily from the distilled training dataset.

\subsection{Baselines}
We select mainstream models in the APR fields as the baselines for our experiments. In the APR field, we selected 7 APR models based on traditional machine learning(ML) and 6 APR models based on {large language model(LLM)}. Traditional machine learning APR models include: SequenceR \cite{chen2019sequencer}, CoConut \cite{lutellier2020coconut}, Rewardrepair \cite{ye2022neural}, Recoder \cite{zhu2021syntax}, AlphaRepair \cite{xia2022less}, TENURE \cite{meng2023template}. LLM-based APR models include: Incoder-1B \cite{fried2022incoder}, CodeGen2-1B \cite{nijkamp2023codegen2}, ChatGPT \cite{chatgpt}, ChatRepair \cite{xia2024automated}, ThinkRepair-Codellama \cite{yin2024thinkrepair}, and ThinkRepair-ChatGPT \cite{yin2024thinkrepair}. To ensure the fairness of the experimental results, {we use the parameter efficient fine-tuning tool LoRA \cite{hu2021lora} to fine-tune Incoder-1B and CodeGen2-1B on the NARRepair training dataset.} These baseline models include models with simple structure and fast {repair} speed like SequenceR, and models with complex structure, good performance but slow {repair} speed like AlphaRepair. In addition, we also select ThinkRepair-ChatGPT, which (to our knowledge) is the state-of-the-art models on the Defect4J and QuixBugs datasets.  {For the repair accuracy of the baselines, we directly adopt the results published in the corresponding papers under the no time limit scenario to promote fair comparison. In the time limit scenario, we use the results obtained by re-running the baseline models. When re-running, we only added time limits to the baseline models, so we cannot obtain results on datasets that the baseline models do not cover.} For example, since SequenceR only publishes the experimental results for the Defect4J v1.2 dataset under the perfect defect location assumption, we can only obtain results on Defects4J v1.2 and consider the results for other datasets as unknown. We refer to the experimental settings of ThinkRepair\cite{yin2024thinkrepair} and conduct experiments in the single-function bug fixing scenario, \emph{i.e.}, the fix is located in a single function. Note that both the single-hunk fixes and single-line fixes are the subsets of the single-function fixes.
For the repair speed of the baselines, we re-run the models provided by the authors to establish the needed time. Please note that our time includes patch generation time and patch verification time. 
For all baseline models in the experiment, we similarly use the ExpressAPR tool \cite{xiao2023expressapr} for patch verification in order to enable fair comparison with NARRepair.

\subsection{Implementation Details}
We use Pytorch \cite{pytorch} to implement the NARRepair model. During training, same as in previous work, we use the Adam optimizer \cite{kingma2014adam} to update the model parameters. During the optimization process, to feed the model as much data as possible, we set the batch size and epoch in all our experiments to be 50 and 200 respectively. As the training process proceeds, the learning rate is adjusted (ranging from 0 to 0.00005) to adapt to the learning situation at different stages of the model. The maximum sequence length is set to be 1024, and the token out of range is ignored. After experimental verification, we set the maximum repair length to be 100. In terms of equipment, our evaluations are performed on an Ubuntu 22.04.5 server equipped with two RTX A6000 GPUs.

\section{Evaluation Results} \label{evaluationresult}
This section introduces how each research question is pursued in detail and the corresponding results.

\subsection{(RQ1) Results of Comparison with the Baselines}
{To simulate the urgency of fixing bugs, we set multiple time limits for the APR model to fix a single bug.} Zhang et al. \cite{zhang2022program} studied the time cost of manually fixing bugs, and {the fastest time for developers to fix a bug is 5 minutes according to their results.} {Thus, based on this research result, we set 3-minutes, 5-minutes, and 10-minutes as three time limits to determine whether the APR models can replace manual repair of urgent bugs.} Note that to ensure the completeness of the repair process, the time we calculate includes both patch generation and patch verification. Furthermore, the patches for all models in our experiments were verified using the ExpressAPR tool \cite{xiao2023expressapr}, which can speed up the patch verification process.

\textbf{Results with Perfect Fault Localization.} 
We first compare the repair performance of the NARRepair model with that of the baseline models under the perfect fault localization scenario. The results are presented in Table \ref{tab:t1}. According to the table data, NARRepair fixes more bugs than all APR baselines under the three time constraints. Compared to the best baseline ThinkRepair-ChatGPT, NARRepair shows notable improvements. For the Defects4J v1.2 dataset, NARRepair fixes 9, 15, and 7 more bugs under the 3-minutes, 5-minutes, and 10-minutes limits, respectively. For the Defects4J v2.0 dataset, NARRepair fixes 7, 11, and 5 more bugs under the 3-minutes, 5-minutes, and 10-minutes limits, respectively. For the QuixBugs dataset, NARRepair fixes 4, 5, and 3 more bugs under the 3-minutes, 5-minutes, and 10-minutes limits, respectively. Note that the NARRepair model has only 520M parameters, which is much smaller than the LLM-based APR models. These results show that due to the AR reasoning method and the parameter size of the baseline models (especially the LLM-based APR model with 1B to 7B parameters), the number of bugs they can fix in a limited time is greatly reduced. 
It can also be seen from the table that even without time limit, NARRepair can still maintain similar or even better performance than the baseline models based on traditional machine learning. Overall, these experimental results show that NARRepair is the state-of-the-art APR method in terms of comprehensive indicator that accounts for both repair speed and repair performance. Besides, the results also imply that compared with other APR models, NARRepair is best suited for replacing programmers to fix urgent bugs. 

\begin{table*}[]
\centering
\caption{\label{tab:t1}The Number of Corrected Bugs with the Perfect Fault Information.}
\resizebox{\textwidth}{!}{
\begin{tabular}{llcccccc}
\hline
\multicolumn{2}{c}{\multirow{3}{*}{Models}}                                             & \multicolumn{6}{c}{Time Limit}                                                                                                                                                                                                                                                                          \\ \cline{3-8} 
\multicolumn{2}{c}{}                                                                    & \multicolumn{3}{c}{\textless 3-minutes}                                                                                             & \multicolumn{3}{c}{\textless 5-minutes}                                                                                             \\ \cline{3-8} 
\multicolumn{2}{c}{}                                                                    & \begin{tabular}[c]{@{}c@{}}Defects4J\\ v1.2\end{tabular} & \begin{tabular}[c]{@{}c@{}}Defects4J\\ v2.0\end{tabular} & \multicolumn{1}{l}{QuixBugs} & \begin{tabular}[c]{@{}c@{}}Defects4J\\ v1.2\end{tabular} & \begin{tabular}[c]{@{}c@{}}Defects4J\\ v2.0\end{tabular} & \multicolumn{1}{l}{QuixBugs} \\ \hline
\multirow{6}{*}{ML}  & SequenceR                                                        & 5                                                        & -                                                        & -                            & 10                                                       & -                                                        & -                            \\ \cline{2-8} 
                     & CoCoNut                                                          & 9                                                        & -                                                        & 3                            & 18                                                       & -                                                        & 7                            \\ \cline{2-8} 
                     & Rewardrepair                                                     & 13                                                       & 11                                                       & 5                            & 22                                                       & 19                                                       & 11                           \\ \cline{2-8} 
                     & Recoder                                                          & 15                                                       & 6                                                        & 7                            & 26                                                       & 11                                                       & 9                            \\ \cline{2-8} 
                     & AlphaRepair                                                      & 18                                                       & 9                                                        & 8                            & 30                                                       & 13                                                       & 15                           \\ \cline{2-8} 
                     & TENURE                                                           & 16                                                       & 11                                                       & -                            & 27                                                       & 17                                                       & -                            \\ \hline
\multirow{6}{*}{LLM} & InCoder-1B                                                       & 15                                                       & 10                                                       & 8                            & 28                                                       & 18                                                       & 14                           \\ \cline{2-8} 
                     & CodeGen2-1B                                                      & 17                                                       & 12                                                       & 10                           & 30                                                       & 20                                                       & 16                           \\ \cline{2-8} 
                     & ChatGPT                                                          & 14                                                       & 11                                                       & 12                           & 28                                                       & 18                                                       & 17                           \\ \cline{2-8} 
                     & ChatRepair                                                       & 21                                                       & -                                                        & 14                           & 31                                                       & -                                                        & 20                           \\ \cline{2-8} 
                     & \begin{tabular}[c]{@{}l@{}}ThinkRepair-\\ Codellama\end{tabular} & 23                                                       & 18                                                       & 15                           & 33                                                       & 23                                                       & 21                           \\ \cline{2-8} 
                     & \begin{tabular}[c]{@{}l@{}}ThinkRepair-\\ ChatGPT\end{tabular}   & 26                                                       & 22                                                       & 18                           & 36                                                       & 29                                                       & 24                           \\ \hline
our                  & NARRepair                                                        & \textbf{35}                                                       & \textbf{29}                                                       & \textbf{22 }                          & \textbf{51}                                                       & \textbf{40}                                                       & \textbf{29 }                          \\ \hline
\multicolumn{2}{c}{\multirow{2}{*}{Models}}                                             & \multicolumn{3}{c}{\textless 10-minutes}                                                                                            & \multicolumn{3}{c}{without limit}                                                                                                                  \\ \cline{3-8} 
\multicolumn{2}{c}{}                                                                    & \begin{tabular}[c]{@{}c@{}}Defects4J\\ v1.2\end{tabular} & \begin{tabular}[c]{@{}c@{}}Defects4J\\ v2.0\end{tabular} & \multicolumn{1}{l}{QuixBugs} & \begin{tabular}[c]{@{}c@{}}Defects4J\\ v1.2\end{tabular} & \begin{tabular}[c]{@{}c@{}}Defects4J\\ v2.0\end{tabular} & \multicolumn{1}{l}{QuixBugs} \\ \hline
\multirow{6}{*}{ML}  & SequenceR                                                        & 14                                                       & -                                                        & -                            & 14                                                       & -                                                        & -                            \\ \cline{2-8} 
                     & CoCoNut                                                          & 29                                                       & -                                                        & 11                           & 38                                                       & -                                                        & 13                           \\ \cline{2-8} 
                     & Rewardrepair                                                     & 34                                                       & 30                                                       & 16                           & 48                                                       & 44                                                       & 20                           \\ \cline{2-8} 
                     & Recoder                                                          & 40                                                       & 15                                                       & 13                           & 64                                                       & 19                                                       & 17                           \\ \cline{2-8} 
                     & AlphaRepair                                                      & 46                                                       & 26                                                       & 21                           & 67                                                       & 36                                                       & 28                           \\ \cline{2-8} 
                     & TENURE                                                           & 42                                                       & 30                                                       & -                            & 61                                                       & 43                                                       & -                            \\ \hline
\multirow{6}{*}{LLM} & InCoder-1B                                                       & 41                                                       & 28                                                       & 20                           & 65                                                       & 40                                                       & 24                           \\ \cline{2-8} 
                     & CodeGen2-1B                                                      & 44                                                       & 32                                                       & 22                           & 68                                                       & 42                                                       & 27                           \\ \cline{2-8} 
                     & ChatGPT                                                          & 40                                                       & 35                                                       & 28                           & 71                                                       & 46                                                       & 38                           \\ \cline{2-8} 
                     & ChatRepair                                                       & 48                                                       & -                                                        & 31                           & 76                                                       & -                                                        & \textbf{39}                           \\ \cline{2-8} 
                     & \begin{tabular}[c]{@{}l@{}}ThinkRepair-\\ Codellama\end{tabular} & 50                                                       & 44                                                       & 29                           & 70                                                       & 72                                                       & 38                           \\ \cline{2-8} 
                     & \begin{tabular}[c]{@{}l@{}}ThinkRepair-\\ ChatGPT\end{tabular}   & 67                                                       & 61                                                       & 31                           & \textbf{98  }                                                     & \textbf{107}                                                      & \textbf{39}                           \\ \hline
our                  & NARRepair                                                        & \textbf{74}                                                       & \textbf{66 }                                                      & \textbf{34}                           & 78                                                       & 69                                                       & 36                           \\ \hline
\end{tabular}
}
\end{table*}

\textbf{Results without Perfect Fault Localization.} 
We also compare the performance of the NARRepair model with that of the baseline models when the defect location is not known in advance. For this, we use Ochiai \cite{abreu2007accuracy}, a widely used spectrum-based fault localization tool to establish the suspiciousness scores of buggy statements and rank them accordingly. Since the Ochiai tool cannot perfectly locate all bugs, the number of fixed bugs by all APR models will decrease. The detailed results are also shown in Table \ref{tab:t7}. Under this scenario, NARRepair still outperforms all APR baselines under time limits of 3-minutes, 5-minutes and 10-minutes. Compared to the best baseline ThinkRepair-ChatGPT, NARRepair again has obvious better performances. For the Defects4J v1.2 dataset, NARRepair fixes 7, 11, and 9 more bugs under the 3-minutes, 5-minutes, and 10-minutes limits, respectively. For the Defects4J v2.0 dataset, NARRepair fixes 6, 9, and 6 more bugs under the 3-minutes, 5-minutes, and 10-minutes limits, respectively. For the QuixBugs dataset, NARRepair fixes 4, 4, and 2 more bugs under the 3-minutes, 5-minutes, and 10-minutes limits, respectively.
Since imperfect positioning is more realistic, these data further demonstrate that NARRepair is the most advanced APR method in terms of comprehensive indicator that considers both repair speed and repair performance. Again, the results also imply that NARRepair can better handle urgent bugs.

\begin{table*}[]
\centering
\caption{\label{tab:t7}The Number of Corrected Bugs without the Perfect Fault Information.}
\resizebox{\textwidth}{!}{
\begin{tabular}{llcccccc}
\hline
\multicolumn{2}{c}{\multirow{3}{*}{Models}}                                             & \multicolumn{6}{c}{Time Limit}                                                                                                                                                                                                                                                                          \\ \cline{3-8} 
\multicolumn{2}{c}{}                                                                    & \multicolumn{3}{c}{\textless 3-minutes}                                                                                             & \multicolumn{3}{c}{\textless 5-minutes}                                                                                             \\ \cline{3-8} 
\multicolumn{2}{c}{}                                                                    & \begin{tabular}[c]{@{}c@{}}Defects4J\\ v1.2\end{tabular} & \begin{tabular}[c]{@{}c@{}}Defects4J\\ v2.0\end{tabular} & \multicolumn{1}{l}{QuixBugs} & \begin{tabular}[c]{@{}c@{}}Defects4J\\ v1.2\end{tabular} & \begin{tabular}[c]{@{}c@{}}Defects4J\\ v2.0\end{tabular} & \multicolumn{1}{l}{QuixBugs} \\ \hline
\multirow{6}{*}{ML}  & SequenceR                                                        & -                                                        & -                                                        & -                            & -                                                        & -                                                        & -                            \\ \cline{2-8} 
                     & CoCoNut                                                          & -                                                        & -                                                        & -                            & -                                                        & -                                                        & -                            \\ \cline{2-8} 
                     & Rewardrepair                                                     & 8                                                        & 6                                                        & -                            & 12                                                       & 11                                                       & -                            \\ \cline{2-8} 
                     & Recoder                                                          & 10                                                       & 4                                                        & 4                            & 17                                                       & 9                                                        & 7                            \\ \cline{2-8} 
                     & AlphaRepair                                                      & 14                                                       & -                                                        & -                            & 21                                                       & -                                                        & -                            \\ \cline{2-8} 
                     & TENURE                                                           & 16                                                       & 8                                                        & -                            & 22                                                       & 14                                                       & -                            \\ \hline
\multirow{6}{*}{LLM} & InCoder-1B                                                       & 12                                                       & 7                                                        & 6                            & 20                                                       & 15                                                       & 10                           \\ \cline{2-8} 
                     & CodeGen2-1B                                                      & 14                                                       & 9                                                        & 7                            & 24                                                       & 16                                                       & 13                           \\ \cline{2-8} 
                     & ChatGPT                                                          & 12                                                       & 8                                                        & 8                            & 23                                                       & 15                                                       & 15                           \\ \cline{2-8} 
                     & ChatRepair                                                       & 16                                                       & -                                                        & -                            & 27                                                       & -                                                        & -                            \\ \cline{2-8} 
                     & \begin{tabular}[c]{@{}l@{}}ThinkRepair-\\ Codellama\end{tabular} & 19                                                       & 13                                                       & 11                           & 29                                                       & 20                                                       & 18                           \\ \cline{2-8} 
                     & \begin{tabular}[c]{@{}l@{}}ThinkRepair-\\ ChatGPT\end{tabular}   & 22                                                       & 16                                                       & 13                           & 32                                                       & 24                                                       & 21                           \\ \hline
our                  & NARRepair                                                        & \textbf{29}                                                       & \textbf{22 }                                                      & \textbf{17}                           & \textbf{43  }                                                     & \textbf{33}                                                       & \textbf{25}                           \\ \hline
\multicolumn{2}{c}{\multirow{2}{*}{Models}}                                             & \multicolumn{3}{c}{\textless 10-minutes}                                                                                            & \multicolumn{3}{c}{without limit}                                                                                                                  \\ \cline{3-8} 
\multicolumn{2}{c}{}                                                                    & \begin{tabular}[c]{@{}c@{}}Defects4J\\ v1.2\end{tabular} & \begin{tabular}[c]{@{}c@{}}Defects4J\\ v2.0\end{tabular} & \multicolumn{1}{l}{QuixBugs} & \begin{tabular}[c]{@{}c@{}}Defects4J\\ v1.2\end{tabular} & \begin{tabular}[c]{@{}c@{}}Defects4J\\ v2.0\end{tabular} & \multicolumn{1}{l}{QuixBugs} \\ \hline
\multirow{6}{*}{ML}  & SequenceR                                                        & -                                                        & -                                                        & -                            & -                                                        & -                                                        & -                            \\ \cline{2-8} 
                     & CoCoNut                                                          & -                                                        & -                                                        & -                            & -                                                        & -                                                        & -                            \\ \cline{2-8} 
                     & Rewardrepair                                                     & 23                                                       & 19                                                       & -                            & 27                                                       & 24                                                       & -                            \\ \cline{2-8} 
                     & Recoder                                                          & 31                                                       & 12                                                       & 11                           & 49                                                       & 19                                                       & 17                           \\ \cline{2-8} 
                     & AlphaRepair                                                      & 36                                                       & -                                                        & -                            & 50                                                       & -                                                        & -                            \\ \cline{2-8} 
                     & TENURE                                                           & 38                                                       & 21                                                       & -                            & 52                                                       & 32                                                       & -                            \\ \hline
\multirow{6}{*}{LLM} & InCoder-1B                                                       & 35                                                       & 23                                                       & 15                           & 54                                                       & 34                                                       & 19                           \\ \cline{2-8} 
                     & CodeGen2-1B                                                      & 39                                                       & 25                                                       & 17                           & 55                                                       & 34                                                       & 20                           \\ \cline{2-8} 
                     & ChatGPT                                                          & 36                                                       & 28                                                       & 21                           & 51                                                       & 38                                                       & 27                           \\ \cline{2-8} 
                     & ChatRepair                                                       & 43                                                       & -                                                        & -                            & 68                                                       & -                                                        & -                            \\ \cline{2-8} 
                     & \begin{tabular}[c]{@{}l@{}}ThinkRepair-\\ Codellama\end{tabular} & 47                                                       & 35                                                       & 23                           & 59                                                       & 53                                                       & 25                           \\ \cline{2-8} 
                     & \begin{tabular}[c]{@{}l@{}}ThinkRepair-\\ ChatGPT\end{tabular}   & 55                                                       & 42                                                       & 26                           & \textbf{74}                                                       & \textbf{75}                                                       & \textbf{30  }                         \\ \hline
our                  & NARRepair                                                        & \textbf{64}                                                       & \textbf{48 }                                                      & \textbf{28}                           & 67                                                       & 52                                                       & 28                           \\ \hline
\end{tabular}
}
\end{table*}

\textbf{Results of Model Repair Speed.}
We evaluate the {repair} time of the NARRepair model and the baseline models in GPU environment. When calculating the {repair} time, we refer to previous works \cite{ye2022neural,xia2024automated,xia2023automated} and let each of the baseline models and the NARRepair model generate 200 patches and calculate time to generate and verify all patches. We set a fixed number of interactions for ChatGPT, ChatRepair, and ThinkRepair to generate 200 patches. In the experiment, we use the {repair} time of ML-based APR model SequenceR, which has the fastest {repair} speed, as the baseline and show the results in Table \ref{tab:t2}. Compared with ML-based APR models with complex structure, such as TENURE, the {repair} speed of NARRepair is increased by 3.8 times. Compared to ML-based APR models with simple structure, such as SequenceR, the {repair} speed of NARRepair is still increased by 1.4 times. Compared to the LLM-based APR model, the overall speed of NARRepair is improved by 2.8 to 6.4 times. The data in the table show that although LLMs with larger number of parameters can deliver better performance, they require longer {repair} time. Overall, the results show that compared to other ML-based APR models and LLM-based APR models, the {repair} speed of the NARRrepair model has been greatly improved. 

\begin{table*}[]
\centering
\caption{\label{tab:t2}The Repair Time of the APR Models.}
\resizebox{\textwidth}{!}{
\begin{tabular}{ccccccc}
\hline
Model              & SequenceR   & CoCoNut    & Rewardrepair & Recoder    & AlphaRepair                                                      & TENURE                                                         \\ \hline
Latency on TestSet & 1.4x        & 1.7x       & 2.2x         & 3.6x       & 3.3x                                                             & 3.8x                                                           \\ \hline
Model              & CodeGen2-1B & InCoder-1B & ChatGPT      & ChatRepair & \begin{tabular}[c]{@{}c@{}}ThinkRepair-\\ Codellama\end{tabular} & \begin{tabular}[c]{@{}c@{}}ThinkRepair-\\ ChatGPT\end{tabular} \\ \hline
Latency on TestSet & 2.8x        & 3.4x       & 5.8x         & 6.4x       & 5.0x                                                             & 6.4x                                                           \\ \hline
\end{tabular}
}
\end{table*}

\begin{table*}[]
\centering
\caption{\label{tab:t3}The Parameters of the APR models.}
\resizebox{\textwidth}{!}{
\begin{tabular}{cccccccc}
\hline
Model      & Rewardrepair & AlphaRepair & TENURE & CodeGen2 & InCoder & \begin{tabular}[c]{@{}c@{}}ThinkRepair-\\ Codellama\end{tabular} & NARRepair \\ \hline
Parameters & 160M         & 180M        & 120M   & 1000M    & 1000M   & 7000M                                                            & 520M      \\ \hline
\end{tabular}
}
\end{table*}

We also consider the impact of the number of parameters of the APR model on the {repair} speed. Generally speaking, the smaller the number of model parameters, the fewer calculations required during the inference process and the faster the inference speed. To verify that the inference speed of the NARRepair model has nothing to do with the number of parameters of the model, we count the number of parameters of several representative APR baseline models and show them in Table \ref{tab:t3}. From the data in Table \ref{tab:t3}, we can find that the parameters of the NARRepair model are slightly higher than those of the traditional machine learning-based APR model but much lower than those of the LLM-based APR model. This suggests that the acceleration of the inference speed of the NARRepair model is not due to the reduction in the number of parameters.

\begin{tcolorbox}[width=\linewidth,boxrule=0pt,top=1pt, bottom=1pt, left=1pt,right=1pt, colback=gray!20,colframe=gray!20]
\textbf{Answer to RQ1:} The experimental results demonstrate that NARRepair is the most advanced APR method in terms of comprehensive indicator that accounts for both repair speed and repair performance. The experimental results also imply that NARRepair is the most suitable APR model for handling urgent bugs.
\end{tcolorbox}
\begin{table}
\centering
\caption{\label{tab:t4}The Ablation Study on the Defect4J v1.2 Dataset.}
\resizebox{0.7\textwidth}{!}{
\begin{tabular}{lc}
\hline
Model                              & \multicolumn{1}{l}{Perfect FL} \\ \hline
NARRepair                          & \textbf{78}                    \\
--Repair Action Predictor          & 69                             \\
--Inter-token Dependency Extractor & 71                             \\
--Two-stage Decoder                & 61                             \\ \hline
\end{tabular}
}
\end{table}
\subsection{(RQ2) Results of Ablation Study}
To evaluate the contribution of each part of the NARRepair model, we perform an ablation study on the Defect4J v1.2 dataset under the perfect fault localization scenario. Starting from the complete model, we remove specific parts of the model structure separately and observe the impact of the removal on the results. More specifically, we (1) first remove the repair action predictor and instead pass only the repair length to the decoder to observe the impact of repair actions on the results; (2) then remove the inter-token dependency extractor to observe the impact of inter-token dependency information on the results; (3) finally replace the two-stage decoder with a normal NAR decoder to observe the impact of contextual information on the results. The results are shown in Table \ref{tab:t4}.

From the table results, we draw the following conclusions. First, after removing the repair action predictor, the performance of the model drops by 9 under the perfect fault localization scenario. This implies that the prediction of repair actions can effectively avoid the problem of modifying correct tokens into wrong ones. Second, after removing the inter-token dependency extractor, the number of repaired programs by the model drops by 7 under the perfect fault localization scenario. This result verifies that the obtained inter-token dependency information can effectively improve the prediction accuracy of the model. Finally, after removing the two-stage decoder, the number of repaired programs by the model drops by 17 under the perfect fault localization scenario. This shows that the contextual information obtained by the decoder through the \texttt{[Mask]} tag is of great significance for the result. 

\begin{tcolorbox}[width=\linewidth,boxrule=0pt,top=1pt, bottom=1pt, left=1pt,right=1pt, colback=gray!20,colframe=gray!20]
\textbf{Answer to RQ2:} The performance of NARRepair after removing different components shows that: all major components of the proposed method contribute positively to the final results.
\end{tcolorbox}

\subsection{(RQ3) Results of Predicting the Repair Actions and Lengths}
The repair actions and repair length of bugs directly affect the final output results of the NARRepair model. Therefore, we further analyze the performance of NARRepair in predicting the repair actions and repair lengths. In the experiment, we divided the Defects4j v1.2 dataset into four categories according to the number of tokens that need to be fixed in the bug: $N \leq 10 $, $10 < N \leq 20$, $20 < N \leq 50$, and $N>50$. \emph{N} means the number of tokens in the bug lines. The four categories of bugs contain 114, 80, 79, and 122 bugs, respectively. We calculate the accuracy of NARRepair in predicting the bug repair actions and repair lengths. Table \ref{tab:t8} shows the performance of NARRepair in predicting repair actions and repair lengths.
\begin{table}[]
\centering
\caption{\label{tab:t8}The Accuracy of Predicting the Repair Actions and Repair Length.}
\resizebox{0.75\textwidth}{!}{
\begin{tabular}{ccc}
\hline
                                                          & Repair Action & Repair Length \\ \hline
$N\leq10$ (114)                              & 83.4\%        & 81.2\%        \\ \hline
$10\textless N \leq20 (80)$ & \textbf{87.3\%}        & \textbf{83.6\% }       \\ \hline
$20\textless N \leq50 (79) $& 80.7\%        & 79.5\%        \\ \hline
$N\textgreater 50 (122)  $                & 79.1\%        & 75.3\%        \\ \hline
Average                                                   & 82.6\%        & 79.8\%        \\ \hline
\end{tabular}
}
\end{table}

From the data in the Table \ref{tab:t8}, we can see that the average accuracy of NARRepair for predicting the repair actions and repair length is 82.6\% and 79.8\% respectively. As the code length increases, the accuracy of NARRepair in predicting the repair action and repair length first increases and then decreases. {In terms of repair actions, NARRepair predicts bugs with length $10 < N \leq 20$ most accurately (87.3\%). In terms of repair lengths, NARRepair also predicts bugs with length $10 < N \leq 20$ most accurately (83.6\%). These results illustrate that NARRepair can accurately predict the repair actions and lengths of bugs.}

\begin{tcolorbox}[width=\linewidth,boxrule=0pt,top=1pt, bottom=1pt, left=1pt,right=1pt, colback=gray!20,colframe=gray!20]
\textbf{Answer to RQ3:} The experimental results show that NARRepair can accurately predict the repair action and repair length of the bugs.
\end{tcolorbox}
 \subsection{(RQ4) Results of Alleviating the Over-Correction Problem}
In Section~\ref{method}, we mentioned that one of the main purposes of the repair action predictor and the two-stage decoder is to avoid changing correct tokens into wrong ones during the repair process. The repair action predictor avoids modifying correct tokens by predicting that the repair action for those tokens is ``Keep''(\ref{action}). The two-stage decoder decodes tokens with low confidence again to avoid some correct tokens from being modified (\ref{decoder}). To explore the effectiveness of our idea in more detail, we respectively remove the repair action predictor module and the two-stage decoder module, and observe the corresponding changes in the output results of the NARRepair model. Among all patches generated by the NARRepair model on the Defect4j v1.2 dataset, we count the average number of correct tokens changed into incorrect ones and present the results in Table \ref{tab:t6}.

From the results in Table \ref{tab:t6}, we can observe that both the repair action predictor module and the two-stage decoder module reduce the number of correct tokens in the code that are modified into incorrect ones. When we remove the repair action predictor from the NARRepair model, the number of correct tokens modified in the patches (generated by the model) increases by 0.9 on average. When we remove the two-stage decoder at the same time, the number of correct tokens modified in the patches (generated by the NARRepair model) further increases by 2.2 on average. This suggests that the two-stage decoder is more effective in avoiding correct tokens wrongly modified into incorrect ones. However, this does not imply that the repair action predictor plays an unimportant role in the entire model. Through the ablation study, we have shown the importance of the repair action predictor. Overall, the result suggests that NARRepair can effectively prevent correct tokens from being modified during the model repair process.
\begin{tcolorbox}[width=\linewidth,boxrule=0pt,top=1pt, bottom=1pt, left=1pt,right=1pt, colback=gray!20,colframe=gray!20]
\textbf{Answer to RQ4:} The experimental results show that the repair action predictor and two-stage decoder in NARRepair effectively reduce the number of over-correction tokens.
\end{tcolorbox}
\begin{table}[]
\centering
\caption{\label{tab:t6}The Number of Correct Tokens Modified.}
\resizebox{0.50\textwidth}{!}{
\begin{tabular}{ll}
\hline
Model                     & Count \\ \hline
NARRepair                 & \textbf{2.2}   \\
--Repair Action Predictor & 3.1   \\
--Two-stage Decoder       & 5.3   \\ \hline
\end{tabular}
}
\end{table}
\begin{figure}
\setlength{\belowcaptionskip}{-15pt}
\centering
\includegraphics[width=0.85\linewidth]{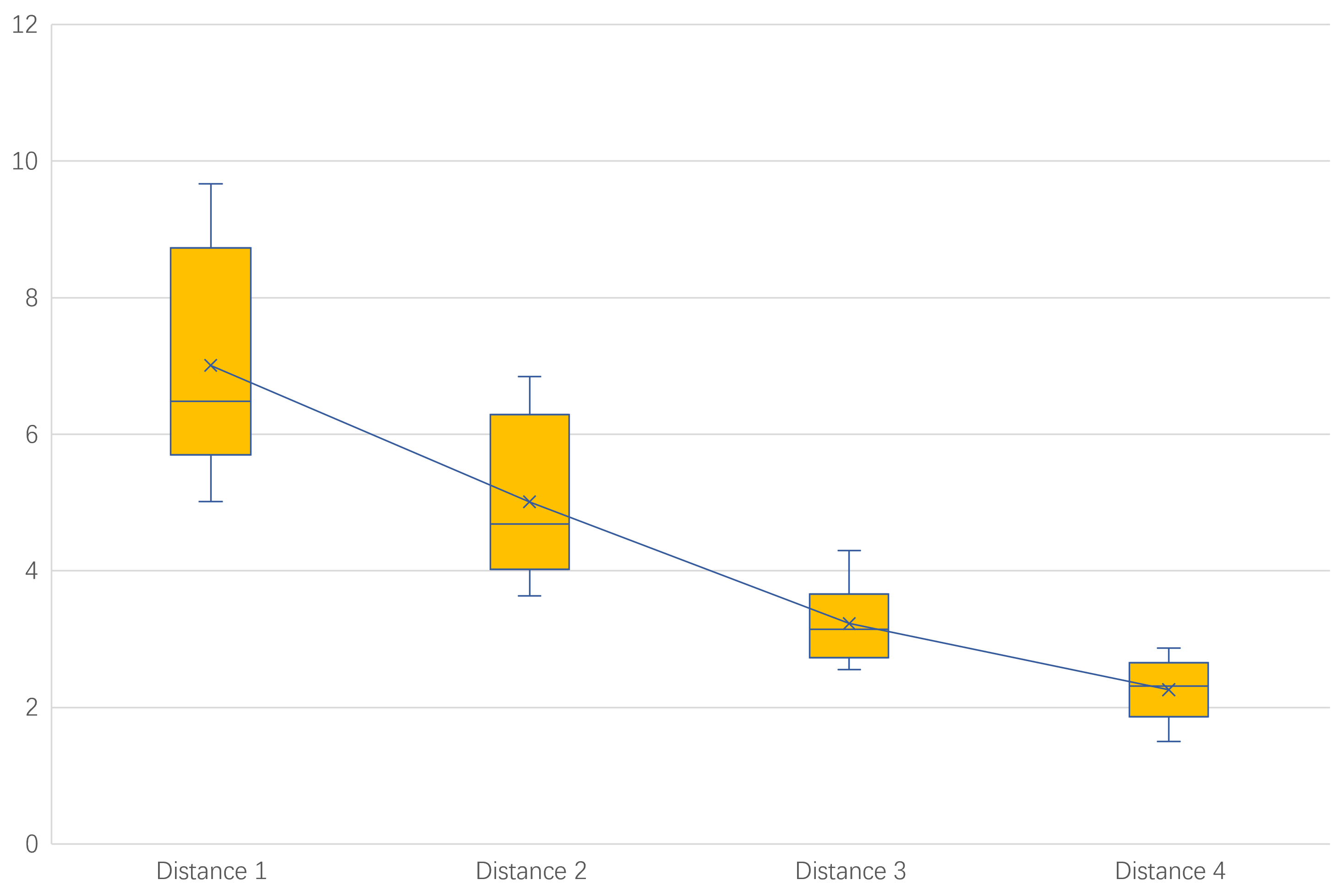}
\caption{\label{fig:frog9}The Cosine Similarity between Nodes in the Abstract Syntax Tree under Different Distance Values. }
\end{figure}

\subsection{(RQ5) Results of the Similarity Between AST Nodes}
To show the degree of correlation between parent nodes and child nodes in the AST, we introduce cosine similarity. Usually, when the cosine similarity of two feature vectors is high, it means that the two vectors are strongly related, and vice versa. We obtain the inter-token dependency matrix output by the trained NARRepair model. Each content in the dependency matrix represents a parent node in the AST. We calculate the feature cosine similarity between the feature vector of the parent node in the dependency matrix and that of each token in the code text (generated by the inter-token dependency extractor). We classify the calculated values by the distance from parent nodes to child nodes in the AST, and the results are shown in Figure \ref{fig:frog9}. According to Figure \ref{fig:frog9}, the similarity between the parent node and the child node gradually decreases as the distance increases. We also find that the farther the distance between the parent node and the child node, the greater the fluctuation range of similarity. 

\begin{figure}
\setlength{\belowcaptionskip}{-15pt}
\centering
\includegraphics[width=0.95\linewidth]{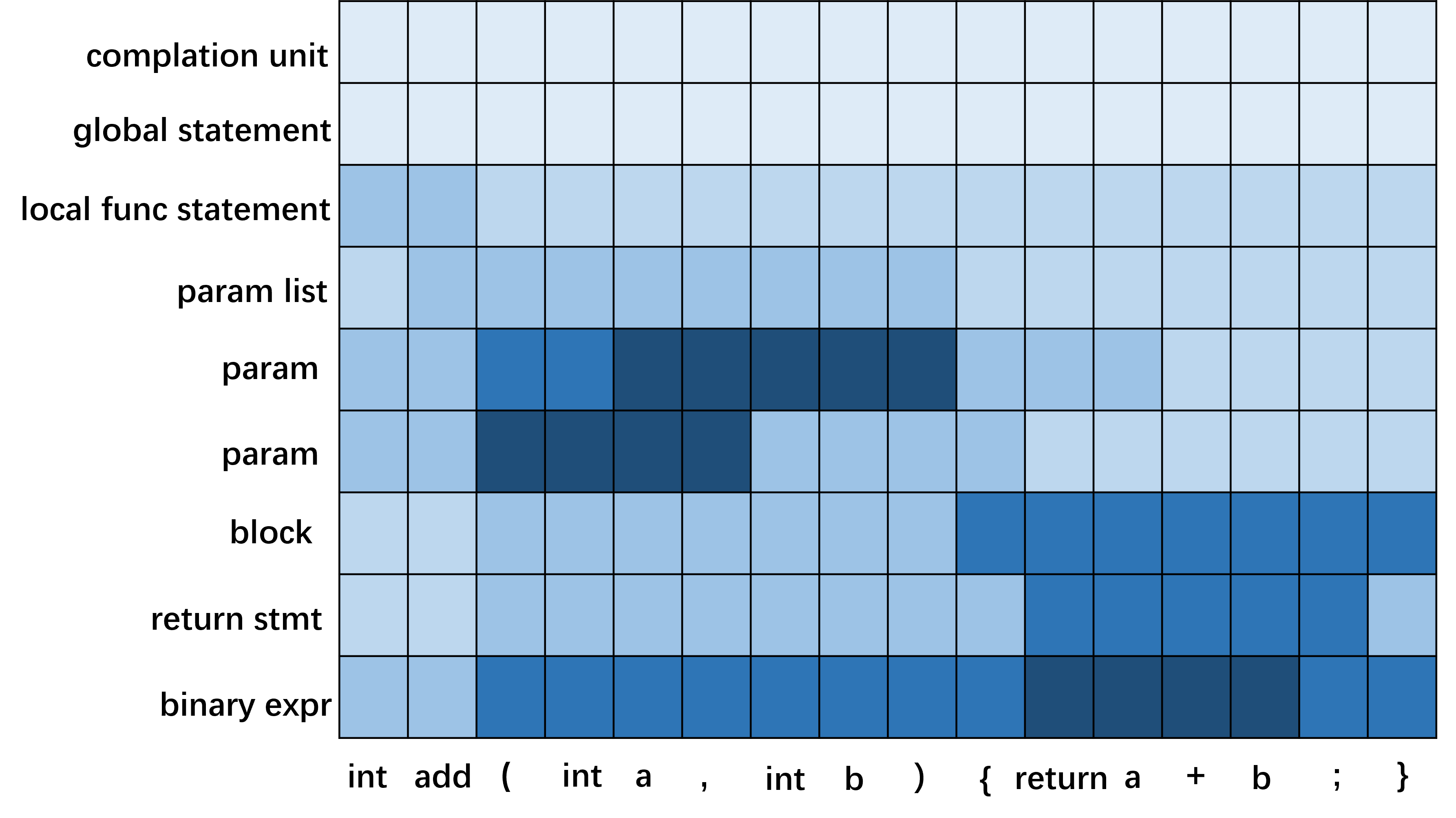}
\caption{\label{fig:frog10}The Heat Map of the Cosine Similarity between Nodes in the AST of the Code Text \texttt{int add (int a, int b) \{return a+b;\}.}}
\end{figure}

To show the relationship between parent nodes and child nodes more concretely, we take the code text \texttt{int add (int a, int b) \{return a+b;\}} as an example and calculate the cosine similarity between the parent node and the child nodes in its AST. The calculated results are shown in Figure \ref{fig:frog10} as a heat map. In the heat map, the darker the color of the square, the higher the cosine similarity between the vectors and the closer the relationship between the two vectors. From Figure \ref{fig:frog10}, we can see that the cosine similarities between the parent node \texttt{binary expr} and its child nodes \texttt{a}, \texttt{+}, and \texttt{b} are all high, and the similarity between it and other tokens decreases as the distance increases. The nodes \texttt{int a} and \texttt{int b} and their parent node \texttt{param} also comply with this trend. These results verify our hypothesis that the relationship between parent nodes and child nodes is close and the closeness decreases with the increase of distance. In addition, this result also justifies the rationality of our method of using the nearest common parent node as the dependency relationship between tokens.
\begin{tcolorbox}[width=\linewidth,boxrule=0pt,top=1pt, bottom=1pt, left=1pt,right=1pt, colback=gray!20,colframe=gray!20]
\textbf{Answer to RQ5:} The experimental results show that the nearest common parent node between two nodes in AST can effectively represent the relationship between nodes.
\end{tcolorbox}

\subsection{Threats to Validity}
 Our results should be interpreted with several threats to validity in mind, and we here discuss them.

\textbf{Internal Validity.} Threats to internal validity might come from the potential faults in the implementations of NARRepair itself and its evaluation. To avoid faults in the implementation of NARRepair, our implementation is mainly based on mature machine learning and program analysis libraries, such as 
Fairseq \cite{ott-etal-2019-fairseq} and Tree-Sitter \cite{tree-sitter}. In addition, we have performed thorough testing to ensure the correctness of NARRepair. To alleviate the threats to evaluation, we use the reported results published in the respective papers as baselines to facilitate fair comparison. When running the models for the baselines is necessary, we re-run the models provided by the authors and strictly follow the given guidelines. Furthermore, note that the whole artifact for this article is made available online for scrutiny. 

\textbf{External Validity.} A potential threat to external validity concerns the representativeness of the benchmark used in our experiment. To mitigate this threat as much as possible, 
we used three widely used datasets in the APR community for evaluating NARRepair, including Defects4J v1.2, Defects4J v2.0, and QuixBugs.
The obtained results illustrate the effectiveness and generalizability of NARRepair. However, we believe that additional evaluations on other benchmarks (such as \cite{9401985}) can further confirm its effectiveness and generalizability. In addition, since the used three datasets contain only Java bugs, further studies are needed to investigate the effectiveness of NARRepair on other programming languages.


\section{Conclusions and Future Work}\label{conclusion}
In this paper, to increase the {repair} speed while maintaining the accuracy of repairing buggy code, we propose NARRepair, a non-autoregressive model for automatic program repair. To solve the issues of wrongly modifying correct tokens into wrong ones, missing inter-token dependency information, and missing contextual information that are generally associated with non-autoregressive models, we propose a repair action predictor, inter-token dependency extractor, and two-stage decoder in NARRepair for addressing the three issues respectively. We evaluate the performance of the NARRepair model on three widely used datasets for automatic program repair tasks. The results show that NARRepair outperforms other APR models within the time limit, proving that NARRepair is more suitable for handling urgent bugs. In addition, while maintaining high repair accuracy, the {repair} speed of NARRepair is 1.4 to 6.4 times faster than other APR models. 

For future work, we will focus particularly on achieving higher accuracy while increasing the inference speed, even better than that of the state-of-the-art AR models. In addition, we will evaluate the performance of NARRepair using more benchmarks and additional programming languages to verify the effectiveness and generalizability of the model. Finally, we also plan to apply our key ideas to other software engineering tasks for which the inference speed is of vital importance. Indeed, we hope that our work can arouse the interest of software engineering researchers in the inference speed of deep learning models for software engineering tasks, which can make software engineering research better meet actual developer needs.
\bibliographystyle{IEEEtran}
\balance 
\bibliography{sample-base}

\begin{thebibliography}{10}
\providecommand{\url}[1]{#1}
\csname url@samestyle\endcsname
\providecommand{\newblock}{\relax}
\providecommand{\bibinfo}[2]{#2}
\providecommand{\BIBentrySTDinterwordspacing}{\spaceskip=0pt\relax}
\providecommand{\BIBentryALTinterwordstretchfactor}{4}
\providecommand{\BIBentryALTinterwordspacing}{\spaceskip=\fontdimen2\font plus
\BIBentryALTinterwordstretchfactor\fontdimen3\font minus \fontdimen4\font\relax}
\providecommand{\BIBforeignlanguage}[2]{{%
\expandafter\ifx\csname l@#1\endcsname\relax
\typeout{** WARNING: IEEEtran.bst: No hyphenation pattern has been}%
\typeout{** loaded for the language `#1'. Using the pattern for}%
\typeout{** the default language instead.}%
\else
\language=\csname l@#1\endcsname
\fi
#2}}
\providecommand{\BIBdecl}{\relax}
\BIBdecl

\bibitem{le2011genprog}
C.~Le~Goues, T.~Nguyen, S.~Forrest, and W.~Weimer, ``Genprog: A generic method for automatic software repair,'' \emph{Ieee transactions on software engineering}, vol.~38, no.~1, pp. 54--72, 2011.

\bibitem{nguyen2013semfix}
H.~D.~T. Nguyen, D.~Qi, A.~Roychoudhury, and S.~Chandra, ``Semfix: Program repair via semantic analysis,'' in \emph{2013 35th International Conference on Software Engineering (ICSE)}.\hskip 1em plus 0.5em minus 0.4em\relax IEEE, 2013, pp. 772--781.

\bibitem{kim2013automatic}
D.~Kim, J.~Nam, J.~Song, and S.~Kim, ``Automatic patch generation learned from human-written patches,'' in \emph{2013 35th International Conference on Software Engineering (ICSE)}.\hskip 1em plus 0.5em minus 0.4em\relax IEEE, 2013, pp. 802--811.

\bibitem{liu2018mining}
K.~Liu, D.~Kim, T.~F. Bissyand{\'e}, S.~Yoo, and Y.~Le~Traon, ``Mining fix patterns for findbugs violations,'' \emph{IEEE Transactions on Software Engineering}, vol.~47, no.~1, pp. 165--188, 2018.

\bibitem{le2016history}
X.~B.~D. Le, D.~Lo, and C.~Le~Goues, ``History driven program repair,'' in \emph{2016 IEEE 23rd international conference on software analysis, evolution, and reengineering (SANER)}, vol.~1.\hskip 1em plus 0.5em minus 0.4em\relax IEEE, 2016, pp. 213--224.

\bibitem{Prophet}
F.~Long and M.~Rinard, ``Automatic patch generation by learning correct code,'' in \emph{POPL}, 2016, pp. 298--312.

\bibitem{rsrepair}
Y.~Qi, X.~Mao, Y.~Lei, Z.~Dai, and C.~Wang, ``The strength of random search on automated program repair,'' in \emph{ICSE}, 2014, pp. 254--265.

\bibitem{angelix}
S.~Mechtaev, J.~Yi, and A.~Roychoudhury, ``Angelix: Scalable multiline program patch synthesis via symbolic analysis,'' in \emph{ICSE}, 2016, pp. 691--701.

\bibitem{10172854}
\BIBentryALTinterwordspacing
Z.~Fan, X.~Gao, M.~Mirchev, A.~Roychoudhury, and S.~Tan, ``Automated repair of programs from large language models,'' in \emph{2023 IEEE/ACM 45th International Conference on Software Engineering (ICSE)}.\hskip 1em plus 0.5em minus 0.4em\relax Los Alamitos, CA, USA: IEEE Computer Society, may 2023, pp. 1469--1481. [Online]. Available: \url{https://doi.ieeecomputersociety.org/10.1109/ICSE48619.2023.00128}
\BIBentrySTDinterwordspacing

\bibitem{yutse}
Z.~Yu, M.~Martinez, Z.~Chen, T.~F. Bissyandé, and M.~Monperrus, ``Learning the relation between code features and code transforms with structured prediction,'' \emph{IEEE Transactions on Software Engineering}, vol.~49, no.~7, pp. 3872--3900, 2023.

\bibitem{yuemse}
Z.~Yu, M.~Martinez, B.~Danglot, T.~Durieux, and M.~Monperrus, ``Alleviating patch overfitting with automatic test generation: A study of feasibility and effectiveness for the nopol repair system,'' \emph{Empirical Software Engineering}, 05 2018.

\bibitem{yuan2018arja}
Y.~Yuan and W.~Banzhaf, ``Arja: Automated repair of java programs via multi-objective genetic programming,'' \emph{IEEE Transactions on software engineering}, vol.~46, no.~10, pp. 1040--1067, 2018.

\bibitem{mechtaev2015directfix}
S.~Mechtaev, J.~Yi, and A.~Roychoudhury, ``Directfix: Looking for simple program repairs,'' in \emph{2015 IEEE/ACM 37th IEEE International Conference on Software Engineering}, vol.~1.\hskip 1em plus 0.5em minus 0.4em\relax IEEE, 2015, pp. 448--458.

\bibitem{yu2021deeprepair}
B.~Yu, H.~Qi, Q.~Guo, F.~Juefei-Xu, X.~Xie, L.~Ma, and J.~Zhao, ``Deeprepair: Style-guided repairing for deep neural networks in the real-world operational environment,'' \emph{IEEE Transactions on Reliability}, vol.~71, no.~4, pp. 1401--1416, 2021.

\bibitem{chen2019sequencer}
Z.~Chen, S.~Kommrusch, M.~Tufano, L.-N. Pouchet, D.~Poshyvanyk, and M.~Monperrus, ``Sequencer: Sequence-to-sequence learning for end-to-end program repair,'' \emph{IEEE Transactions on Software Engineering}, vol.~47, no.~9, pp. 1943--1959, 2019.

\bibitem{9393494}
B.~Baudry, Z.~Chen, K.~Etemadi, H.~Fu, D.~Ginelli, S.~Kommrusch, M.~Martinez, M.~Monperrus, J.~Ron, H.~Ye, and Z.~Yu, ``A software-repair robot based on continual learning,'' \emph{IEEE Software}, vol.~38, no.~4, pp. 28--35, 2021.

\bibitem{zhu2021syntax}
Q.~Zhu, Z.~Sun, Y.-a. Xiao, W.~Zhang, K.~Yuan, Y.~Xiong, and L.~Zhang, ``A syntax-guided edit decoder for neural program repair,'' in \emph{Proceedings of the 29th ACM Joint Meeting on European Software Engineering Conference and Symposium on the Foundations of Software Engineering}, 2021, pp. 341--353.

\bibitem{CODIT}
S.~Chakraborty, Y.~Ding, M.~Allamanis, and B.~Ray, ``Codit: Code editing with tree-based neural models,'' \emph{IEEE Transactions on Software Engineering}, vol.~48, no.~4, pp. 1385--1399, 2022.

\bibitem{lutellier2020coconut}
T.~Lutellier, H.~V. Pham, L.~Pang, Y.~Li, M.~Wei, and L.~Tan, ``Coconut: combining context-aware neural translation models using ensemble for program repair,'' in \emph{Proceedings of the 29th ACM SIGSOFT international symposium on software testing and analysis}, 2020, pp. 101--114.

\bibitem{yin2024thinkrepair}
X.~Yin, C.~Ni, S.~Wang, Z.~Li, L.~Zeng, and X.~Yang, ``Thinkrepair: Self-directed automated program repair,'' in \emph{Proceedings of the 33rd ACM SIGSOFT International Symposium on Software Testing and Analysis}, 2024, pp. 1274--1286.

\bibitem{10.1145/3510003.3510040}
\BIBentryALTinterwordspacing
Y.~Noller, R.~Shariffdeen, X.~Gao, and A.~Roychoudhury, ``Trust enhancement issues in program repair,'' in \emph{Proceedings of the 44th International Conference on Software Engineering}, ser. ICSE '22.\hskip 1em plus 0.5em minus 0.4em\relax New York, NY, USA: Association for Computing Machinery, 2022, p. 2228–2240. [Online]. Available: \url{https://doi.org/10.1145/3510003.3510040}
\BIBentrySTDinterwordspacing

\bibitem{9609108}
J.~Liang, R.~Ji, J.~Jiang, S.~Zhou, Y.~Lou, Y.~Xiong, and G.~Huang, ``Interactive patch filtering as debugging aid,'' in \emph{2021 IEEE International Conference on Software Maintenance and Evolution (ICSME)}, 2021, pp. 239--250.

\bibitem{xiao2023expressapr}
\BIBentryALTinterwordspacing
Y.-A. Xiao, C.~Yang, B.~Wang, and Y.~Xiong, ``Accelerating patch validation for program repair with interception-based execution scheduling,'' \emph{IEEE Trans. Softw. Eng.}, vol.~50, no.~3, p. 618–635, Mar. 2024. [Online]. Available: \url{https://doi.org/10.1109/TSE.2024.3359969}
\BIBentrySTDinterwordspacing

\bibitem{ahmed2020characterizing}
U.~Z. Ahmed, N.~Srivastava, R.~Sindhgatta, and A.~Karkare, ``Characterizing the pedagogical benefits of adaptive feedback for compilation errors by novice programmers,'' in \emph{Proceedings of the ACM/IEEE 42nd International Conference on Software Engineering: Software Engineering Education and Training}, 2020, pp. 139--150.

\bibitem{gaimon1989real}
C.~Gaimon and G.~L. Thompson, ``A real-time solution for preventive and repair maintenance,'' \emph{Optimal Control Applications and Methods}, vol.~10, no.~3, pp. 211--228, 1989.

\bibitem{steinbauer2005real}
G.~Steinbauer, M.~M{\"o}rth, and F.~Wotawa, ``Real-time diagnosis and repair of faults of robot control software,'' in \emph{Robot Soccer World Cup}.\hskip 1em plus 0.5em minus 0.4em\relax Springer, 2005, pp. 13--23.

\bibitem{nazar2015improving}
G.~L. Nazar, ``Improving fpga repair under real-time constraints,'' \emph{Microelectronics Reliability}, vol.~55, no.~7, pp. 1109--1119, 2015.

\bibitem{gu2018non}
J.~Gu, J.~Bradbury, C.~Xiong, V.~O. Li, and R.~Socher, ``Non-autoregressive neural machine translation,'' in \emph{International Conference on Learning Representations}, 2018.

\bibitem{li2019hint}
Z.~Li, Z.~Lin, D.~He, F.~Tian, T.~Qin, L.~Wang, and T.-Y. Liu, ``Hint-based training for non-autoregressive machine translation,'' in \emph{Proceedings of the 2019 Conference on Empirical Methods in Natural Language Processing and the 9th International Joint Conference on Natural Language Processing (EMNLP-IJCNLP)}, 2019, pp. 5708--5713.

\bibitem{10.1145/3649594}
\BIBentryALTinterwordspacing
F.~Liu, Z.~Fu, G.~Li, Z.~Jin, H.~Liu, Y.~Hao, and L.~Zhang, ``Non-autoregressive line-level code completion,'' \emph{ACM Trans. Softw. Eng. Methodol.}, vol.~33, no.~5, Jun. 2024. [Online]. Available: \url{https://doi.org/10.1145/3649594}
\BIBentrySTDinterwordspacing

\bibitem{just2014defects4j}
R.~Just, D.~Jalali, and M.~D. Ernst, ``Defects4j: A database of existing faults to enable controlled testing studies for java programs,'' in \emph{Proceedings of the 2014 international symposium on software testing and analysis}, 2014, pp. 437--440.

\bibitem{lin2017quixbugs}
D.~Lin, J.~Koppel, A.~Chen, and A.~Solar-Lezama, ``Quixbugs: A multi-lingual program repair benchmark set based on the quixey challenge,'' in \emph{Proceedings Companion of the 2017 ACM SIGPLAN international conference on systems, programming, languages, and applications: software for humanity}, 2017, pp. 55--56.

\bibitem{yangoldNARRepair}
\BIBentryALTinterwordspacing
Z.~Yang, Z.~Yang, and Z.~Yu, ``Narrepair: Non-autoregressive code generation model for automatic program repair,'' 2024. [Online]. Available: \url{https://arxiv.org/abs/2406.16526}
\BIBentrySTDinterwordspacing

\bibitem{debugging}
I.~Vessey, ``Expertise in debugging computer programs: A process analysis,'' \emph{International Journal of Man-Machine Studies}, vol.~23, no.~5, pp. 459--494, 1985.

\bibitem{multiple-fault}
Z.~Yu, C.~Bai, and K.-Y. Cai, ``Does the failing test execute a single or multiple faults? an approach to classifying failing tests,'' in \emph{Proceedings of the 37th International Conference on Software Engineering - Volume 1}, ser. ICSE '15.\hskip 1em plus 0.5em minus 0.4em\relax IEEE Press, 2015, p. 924–935.

\bibitem{10.1145/3611643.3616338}
\BIBentryALTinterwordspacing
Y.~Du and Z.~Yu, ``Pre-training code representation with semantic flow graph for effective bug localization,'' in \emph{Proceedings of the 31st ACM Joint European Software Engineering Conference and Symposium on the Foundations of Software Engineering}, ser. ESEC/FSE 2023.\hskip 1em plus 0.5em minus 0.4em\relax New York, NY, USA: Association for Computing Machinery, 2023, p. 579–591. [Online]. Available: \url{https://doi.org/10.1145/3611643.3616338}
\BIBentrySTDinterwordspacing

\bibitem{urli2018design}
S.~Urli, Z.~Yu, L.~Seinturier, and M.~Monperrus, ``How to design a program repair bot? insights from the repairnator project. in 2018 ieee/acm 40th international conference on software engineering: Software engineering in practice track (icse-seip),'' \emph{IEEE Computer Society, Los Alamitos, CA, USA}, pp. 95--104, 2018.

\bibitem{yuguifl}
Z.~Yu, H.~Hu, C.~Bai, K.-Y. Cai, and W.~E. Wong, ``Gui software fault localization using n-gram analysis,'' in \emph{2011 IEEE 13th International Symposium on High-Assurance Systems Engineering}, 2011, pp. 325--332.

\bibitem{10.1016/j.infsof.2013.07.004}
\BIBentryALTinterwordspacing
Z.~Yu, C.~Bai, and K.-Y. Cai, ``Mutation-oriented test data augmentation for gui software fault localization,'' \emph{Inf. Softw. Technol.}, vol.~55, no.~12, p. 2076–2098, Dec. 2013. [Online]. Available: \url{https://doi.org/10.1016/j.infsof.2013.07.004}
\BIBentrySTDinterwordspacing

\bibitem{jobstmann2005program}
B.~Jobstmann, A.~Griesmayer, and R.~Bloem, ``Program repair as a game,'' in \emph{Computer Aided Verification: 17th International Conference, CAV 2005, Edinburgh, Scotland, UK, July 6-10, 2005. Proceedings 17}.\hskip 1em plus 0.5em minus 0.4em\relax Springer, 2005, pp. 226--238.

\bibitem{xuan2016nopol}
J.~Xuan, M.~Martinez, F.~Demarco, M.~Clement, S.~L. Marcote, T.~Durieux, D.~Le~Berre, and M.~Monperrus, ``Nopol: Automatic repair of conditional statement bugs in java programs,'' \emph{IEEE Transactions on Software Engineering}, vol.~43, no.~1, pp. 34--55, 2016.

\bibitem{wei2010automated}
Y.~Wei, Y.~Pei, C.~A. Furia, L.~S. Silva, S.~Buchholz, B.~Meyer, and A.~Zeller, ``Automated fixing of programs with contracts,'' in \emph{Proceedings of the 19th international symposium on Software testing and analysis}, 2010, pp. 61--72.

\bibitem{liu2019tbar}
K.~Liu, A.~Koyuncu, D.~Kim, and T.~F. Bissyand{\'e}, ``Tbar: Revisiting template-based automated program repair,'' in \emph{Proceedings of the 28th ACM SIGSOFT International Symposium on Software Testing and Analysis}, 2019, pp. 31--42.

\bibitem{long2015staged}
F.~Long and M.~Rinard, ``Staged program repair with condition synthesis,'' in \emph{Proceedings of the 2015 10th Joint Meeting on Foundations of Software Engineering}, 2015, pp. 166--178.

\bibitem{gupta2017deepfix}
R.~Gupta, S.~Pal, A.~Kanade, and S.~Shevade, ``Deepfix: Fixing common c language errors by deep learning,'' in \emph{Proceedings of the aaai conference on artificial intelligence}, vol.~31, no.~1, 2017.

\bibitem{ye2022neural}
H.~Ye, M.~Martinez, and M.~Monperrus, ``Neural program repair with execution-based backpropagation,'' in \emph{Proceedings of the 44th International Conference on Software Engineering}, 2022, pp. 1506--1518.

\bibitem{xia2022less}
C.~S. Xia and L.~Zhang, ``Less training, more repairing please: revisiting automated program repair via zero-shot learning,'' in \emph{Proceedings of the 30th ACM Joint European Software Engineering Conference and Symposium on the Foundations of Software Engineering}, 2022, pp. 959--971.

\bibitem{meng2023template}
X.~Meng, X.~Wang, H.~Zhang, H.~Sun, X.~Liu, and C.~Hu, ``Template-based neural program repair,'' in \emph{2023 IEEE/ACM 45th International Conference on Software Engineering (ICSE)}.\hskip 1em plus 0.5em minus 0.4em\relax IEEE, 2023, pp. 1456--1468.

\bibitem{xia2024automated}
C.~S. Xia and L.~Zhang, ``Automated program repair via conversation: Fixing 162 out of 337 bugs for \$0.42 each using chatgpt,'' in \emph{Proceedings of the 33rd ACM SIGSOFT International Symposium on Software Testing and Analysis}, 2024, pp. 819--831.

\bibitem{shu2020latent}
R.~Shu, J.~Lee, H.~Nakayama, and K.~Cho, ``Latent-variable non-autoregressive neural machine translation with deterministic inference using a delta posterior,'' in \emph{Proceedings of the aaai conference on artificial intelligence}, vol.~34, no.~05, 2020, pp. 8846--8853.

\bibitem{ran2021guiding}
Q.~Ran, Y.~Lin, P.~Li, and J.~Zhou, ``Guiding non-autoregressive neural machine translation decoding with reordering information,'' in \emph{Proceedings of the AAAI Conference on Artificial Intelligence}, vol.~35, no.~15, 2021, pp. 13\,727--13\,735.

\bibitem{ma2019flowseq}
X.~Ma, C.~Zhou, X.~Li, G.~Neubig, and E.~Hovy, ``Flowseq: Non-autoregressive conditional sequence generation with generative flow,'' \emph{arXiv preprint arXiv:1909.02480}, 2019.

\bibitem{stern2019insertion}
M.~Stern, W.~Chan, J.~Kiros, and J.~Uszkoreit, ``Insertion transformer: Flexible sequence generation via insertion operations,'' in \emph{International Conference on Machine Learning}.\hskip 1em plus 0.5em minus 0.4em\relax PMLR, 2019, pp. 5976--5985.

\bibitem{gui2023non}
S.~Gui, C.~Shao, Z.~Ma, Y.~Chen, Y.~Feng \emph{et~al.}, ``Non-autoregressive machine translation with probabilistic context-free grammar,'' \emph{Advances in Neural Information Processing Systems}, vol.~36, pp. 5598--5615, 2023.

\bibitem{bao2023non}
G.~Bao, Z.~Teng, H.~Zhou, J.~Yan, and Y.~Zhang, ``Non-autoregressive document-level machine translation,'' in \emph{Findings of the Association for Computational Linguistics: EMNLP 2023}, 2023, pp. 14\,791--14\,803.

\bibitem{liu2023selective}
M.~Liu, Y.~Bao, C.~Zhao, and S.~Huang, ``Selective knowledge distillation for non-autoregressive neural machine translation,'' in \emph{Proceedings of the AAAI Conference on Artificial Intelligence}, vol.~37, no.~11, 2023, pp. 13\,246--13\,254.

\bibitem{tandiffnorm}
W.~Tan, J.~Zhang, L.~Shen, D.~Khashabi, and P.~Koehn, ``Diffnorm: Self-supervised normalization for non-autoregressive speech-to-speech translation,'' in \emph{The Thirty-eighth Annual Conference on Neural Information Processing Systems}.

\bibitem{vaswani2017attention}
A.~Vaswani, N.~Shazeer, N.~Parmar, J.~Uszkoreit, L.~Jones, A.~N. Gomez, {\L}.~Kaiser, and I.~Polosukhin, ``Attention is all you need,'' \emph{Advances in neural information processing systems}, vol.~30, 2017.

\bibitem{kalchbrenner2014convolutional}
N.~Kalchbrenner, E.~Grefenstette, and P.~Blunsom, ``A convolutional neural network for modelling sentences,'' \emph{arXiv preprint arXiv:1404.2188}, 2014.

\bibitem{tree-sitter}
Tree-sitter. (2023) Https://tree-sitter.github.io/tree-sitter/.

\bibitem{dozat2016deep}
T.~Dozat and C.~D. Manning, ``Deep biaffine attention for neural dependency parsing,'' \emph{arXiv preprint arXiv:1611.01734}, 2016.

\bibitem{devlin2018bert}
J.~Devlin, M.-W. Chang, K.~Lee, and K.~Toutanova, ``Bert: Pre-training of deep bidirectional transformers for language understanding,'' \emph{arXiv preprint arXiv:1810.04805}, 2018.

\bibitem{feng2020codebert}
Z.~Feng, D.~Guo, D.~Tang, N.~Duan, X.~Feng, M.~Gong, L.~Shou, B.~Qin, T.~Liu, D.~Jiang \emph{et~al.}, ``Codebert: A pre-trained model for programming and natural languages,'' \emph{arXiv preprint arXiv:2002.08155}, 2020.

\bibitem{liu2019roberta}
Y.~Liu, M.~Ott, N.~Goyal, J.~Du, M.~Joshi, D.~Chen, O.~Levy, M.~Lewis, L.~Zettlemoyer, and V.~Stoyanov, ``Roberta: A robustly optimized bert pretraining approach,'' \emph{arXiv preprint arXiv:1907.11692}, 2019.

\bibitem{guo2020graphcodebert}
D.~Guo, S.~Ren, S.~Lu, Z.~Feng, D.~Tang, S.~Liu, L.~Zhou, N.~Duan, A.~Svyatkovskiy, S.~Fu \emph{et~al.}, ``Graphcodebert: Pre-training code representations with data flow,'' \emph{arXiv preprint arXiv:2009.08366}, 2020.

\bibitem{mutationtesting}
\BIBentryALTinterwordspacing
M.~Papadakis, M.~Kintis, J.~Zhang, Y.~Jia, Y.~L. Traon, and M.~Harman, ``Chapter six - mutation testing advances: An analysis and survey,'' ser. Advances in Computers, A.~M. Memon, Ed.\hskip 1em plus 0.5em minus 0.4em\relax Elsevier, 2019, vol. 112, pp. 275--378. [Online]. Available: \url{https://www.sciencedirect.com/science/article/pii/S0065245818300305}
\BIBentrySTDinterwordspacing

\bibitem{article}
B.~Danglot, O.~Vera~Pérez, Z.~Yu, M.~Monperrus, and B.~Baudry, ``The emerging field of test amplification: A survey,'' 05 2017.

\bibitem{compileroptimization}
J.~Wu, J.~Zheng, Z.~Yang, and Z.~Yu, ``Compiler optimization testing based on optimization-guided equivalence transformations,'' in \emph{FSE}, 2025.

\bibitem{ye2022selfapr}
H.~Ye, M.~Martinez, X.~Luo, T.~Zhang, and M.~Monperrus, ``Selfapr: Self-supervised program repair with test execution diagnostics,'' in \emph{Proceedings of the 37th IEEE/ACM International Conference on Automated Software Engineering}, 2022, pp. 1--13.

\bibitem{wang2019non}
Y.~Wang, F.~Tian, D.~He, T.~Qin, C.~Zhai, and T.-Y. Liu, ``Non-autoregressive machine translation with auxiliary regularization,'' in \emph{Proceedings of the AAAI conference on artificial intelligence}, vol.~33, no.~01, 2019, pp. 5377--5384.

\bibitem{qian2020glancing}
L.~Qian, H.~Zhou, Y.~Bao, M.~Wang, L.~Qiu, W.~Zhang, Y.~Yu, and L.~Li, ``Glancing transformer for non-autoregressive neural machine translation,'' \emph{arXiv preprint arXiv:2008.07905}, 2020.

\bibitem{fried2022incoder}
D.~Fried, A.~Aghajanyan, J.~Lin, S.~Wang, E.~Wallace, F.~Shi, R.~Zhong, W.-t. Yih, L.~Zettlemoyer, and M.~Lewis, ``Incoder: A generative model for code infilling and synthesis,'' \emph{arXiv preprint arXiv:2204.05999}, 2022.

\bibitem{nijkamp2023codegen2}
E.~Nijkamp, H.~Hayashi, C.~Xiong, S.~Savarese, and Y.~Zhou, ``Codegen2: Lessons for training llms on programming and natural languages,'' \emph{arXiv preprint arXiv:2305.02309}, 2023.

\bibitem{chatgpt}
ChatGPT. (2023) Https://openai.com/blog/chatgpt.

\bibitem{hu2021lora}
E.~J. Hu, Y.~Shen, P.~Wallis, Z.~Allen-Zhu, Y.~Li, S.~Wang, L.~Wang, and W.~Chen, ``Lora: Low-rank adaptation of large language models,'' \emph{arXiv preprint arXiv:2106.09685}, 2021.

\bibitem{pytorch}
PyTorch. (2023) Https://pytorch.org/.

\bibitem{kingma2014adam}
D.~P. Kingma and J.~Ba, ``Adam: A method for stochastic optimization,'' \emph{arXiv preprint arXiv:1412.6980}, 2014.

\bibitem{zhang2022program}
Q.~Zhang, Y.~Zhao, W.~Sun, C.~Fang, Z.~Wang, and L.~Zhang, ``Program repair: Automated vs. manual,'' \emph{arXiv preprint arXiv:2203.05166}, 2022.

\bibitem{abreu2007accuracy}
R.~Abreu, P.~Zoeteweij, and A.~J. Van~Gemund, ``On the accuracy of spectrum-based fault localization,'' in \emph{Testing: Academic and industrial conference practice and research techniques-MUTATION (TAICPART-MUTATION 2007)}.\hskip 1em plus 0.5em minus 0.4em\relax IEEE, 2007, pp. 89--98.

\bibitem{xia2023automated}
C.~S. Xia, Y.~Wei, and L.~Zhang, ``Automated program repair in the era of large pre-trained language models,'' in \emph{2023 IEEE/ACM 45th International Conference on Software Engineering (ICSE)}.\hskip 1em plus 0.5em minus 0.4em\relax IEEE, 2023, pp. 1482--1494.

\bibitem{ott-etal-2019-fairseq}
\BIBentryALTinterwordspacing
M.~Ott, S.~Edunov, A.~Baevski, A.~Fan, S.~Gross, N.~Ng, D.~Grangier, and M.~Auli, ``fairseq: A fast, extensible toolkit for sequence modeling,'' in \emph{Proceedings of the 2019 Conference of the North {A}merican Chapter of the Association for Computational Linguistics (Demonstrations)}, W.~Ammar, A.~Louis, and N.~Mostafazadeh, Eds.\hskip 1em plus 0.5em minus 0.4em\relax Minneapolis, Minnesota: Association for Computational Linguistics, Jun. 2019, pp. 48--53. [Online]. Available: \url{https://aclanthology.org/N19-4009}
\BIBentrySTDinterwordspacing

\bibitem{9401985}
Y.~Jiang, H.~Liu, N.~Niu, L.~Zhang, and Y.~Hu, ``Extracting concise bug-fixing patches from human-written patches in version control systems,'' in \emph{2021 IEEE/ACM 43rd International Conference on Software Engineering (ICSE)}, 2021, pp. 686--698.

\end{thebibliography}


\end{document}